\documentclass[fleqn,usenatbib]{mnras}
 
\usepackage{newtxtext,newtxmath}

\usepackage[T1]{fontenc}
\usepackage{ae,aecompl}

\usepackage{graphicx}	
\usepackage{amsmath}	
\usepackage{amssymb}	
\usepackage{pdflscape}  

\title[White Dwarfs in SDSS DR14]{White dwarf and subdwarf stars in the Sloan Digital Sky Survey Data Release 14}
   \author[Kepler et al.]{S. O. Kepler$^{1}$\thanks{kepler@if.ufrgs.br},
                Ingrid Pelisoli$^{2,1}$,
                Detlev Koester$^{3}$,
                Nicole Reindl$^{4}$,
	            Stephan Geier$^2$, 
\newauthor
                Alejandra D. Romero$^{1}$,
                Gustavo Ourique$^{1}$,
	Cristiane de Paula Oliveira$^1$,
\newauthor
              \& Larissa A. Amaral$^{1}$\\
$^{1}$Instituto de F\'{\i}sica, Universidade Federal do Rio Grande do Sul,
              91501-900  Porto-Alegre, RS, Brazil\\
$^2$Institut f\"{u}r Physik und Astronomie, Universit\"{a}tsstandort Golm, Karl-Liebknecht-Str. 24/25, 14467 Potsdam, Germany\\
$^{3}$Institut f\"ur Theoretische Physik und Astrophysik, Universit\"at Kiel, 24098 Kiel, Germany\\
$^4$Department of Physics and Astronomy, University of Leicester, University Road, Leicester LE1 7RH, UK\\
}

\date{Accepted 2019 April 1. Received 2019 February 14; in original form 2018 October 22}

\pubyear{2019}

\begin{document}
\label{firstpage}
\pagerange{\pageref{firstpage}--\pageref{lastpage}}
\maketitle

\begin{abstract}
White dwarfs carry information on the structure and evolution of the Galaxy, especially through their luminosity function and initial-to-final mass relation. 
Very cool white dwarfs provide insight into the early ages of each population.
Examining the spectra of all stars with $3\sigma$ proper motion in the Sloan Digital Sky Survey Data Release 14, we report the classification 
for 
20\,088
spectroscopically confirmed
white dwarfs, plus 415 hot
subdwarfs, and 311 cataclysmic variables.
We obtain $T_\mathrm{eff}$, $\log~g$ and mass for hydrogen
atmosphere white dwarf stars (DAs),
warm helium atmosphere white dwarfs (DBs), hot subdwarfs (sdBs and sdOs), and estimate photometric $T_\mathrm{eff}$ for white 
dwarf stars with continuum spectra (DCs). 
We find 15\,793 sdAs and 447 dCs between the white dwarf cooling sequence and the main sequence, especially below $T_\mathrm{eff}\simeq 10\,000$~K; most are likely low-mass metal-poor main sequence stars, but some could be the result of interacting binary evolution.
\end{abstract}

\begin{keywords}
white dwarfs -- subdwarfs -- catalogues
\end{keywords}



\section{Introduction}

White dwarf stars are the end state for all stars formed with initial masses below around
7--11.8~$M_{\odot}$, depending on metallicity \citep[e.g][]{Ibeling13,Doherty14,Woosley15,Ramos18},
which translates to more than 97\% of all stars. 
Therefore the properties of the white dwarf population reflect the result of the initial mass function, the star formation rate and
the initial-to-final mass relation, for different
metallicities. White dwarf stars are also possible outcomes of the evolution of multiple systems, with 25--30 per cent of white dwarfs estimated to be the result of
mergers \citep{Toonen17}. White dwarfs with masses lower than 0.3--0.45~$M_{\sun}$ are generally explained as outcomes of close binary evolution \citep{Kilic2007}, given that the single progenitors of such low-mass white dwarfs have main sequence lifetimes exceeding the age of the Universe. The formation
mechanism of the so-called extremely-low mass white dwarfs (ELMs) -- those with masses below $\simeq 0.2-0.3 M_\odot$ \citep[e.g.][and references therein]{Sun18, Calcaferro18} -- is similar to that proposed to explain composite hot subdwarf stars \citep[e.g.][]{Heber16}:
the outer envelope is lost after a common envelope or a stable Roche-lobe overflow phase, leaving the stellar core exposed \citep[e.g.][]{li2018}. Hot subdwarfs result when the envelope is lost after He-burning is triggered
in the core -- hence they lie above the zero-age horizontal branch, whereas an ELM will result if the mass is lost when the core He is in a degenerate state, but He
fusion has not been triggered. ELMs show similar $\log~g$ to subdwarfs, but generally lower temperature ($T_\mathrm{eff} \lesssim 20\,000$~K). 

White dwarfs do not present ongoing core nuclear burning, but residual shell burning may occur depending on the thickness of the hydrogen layer. 
ELMs are believed to show residual burning before reaching the final white dwarf cooling track \citep{Corsico12,Istrate16}. This happens in the pre-ELM phase \citep{Maxted14a, Maxted14b},
which can cause them to show luminosities comparable to main sequence and even horizontal branch stars \citep[e.g.][]{RRLyra026}.

Because the timescales for
gravitational settling are of the order
of a few million years or smaller, the atmospheric composition of white dwarf stars is generally simple, with around 80\%
showing solely H lines (spectral class DA). The remaining are dominated by He lines, when the atmospheric temperature is
sufficient to excite the He atoms. The spectral class is DB if only HeI lines are present, and DO if HeII lines are visible
(typically $T_\mathrm{eff} \gtrsim 40\,000$~K). Very cool white dwarfs ($T_\mathrm{eff} \lesssim 5\,000$~K for H atmosphere, 
$T_\mathrm{eff} \lesssim 11\,000$~K for He atmosphere) show featureless spectra
and are classified as DCs. 
A substantial fraction \citep[20--50 per cent,][]{Zuckerman03,Koester14} of white dwarfs show contamination by metals, which can only be explained by ongoing accretion,
except for very hot objects ($T_\mathrm{eff} \gtrsim 50\,000$~K), where radiative levitation can still play a significant role \citep[e.g][]{Barstow14};
a Z is added to the spectral classification to flag metal pollution. In rare cases, for stars classified as DQs, carbon may be dragged to the surface by convection 
\citep[e.g.][]{Koester82}. Cool DQs show spectra similar to dwarf carbon (dC) stars, which are themselves believed to be one outcome binary evolution
\citep{Whitehouse18}.

In this paper we extend the work of \citet{dr7} and \citet{dr10,dr12}, continuing the search for new spectroscopically confirmed white dwarf and subdwarf stars in the 
data release 14 of the Sloan Digital Sky Survey \citep[SDSS DR14,][]{Abolfathi18}. Spectroscopy allows precise determinations of $T_\mathrm{eff}$, $\log~g$, and abundances, serving
as a valuable resource for studying stellar formation and evolution in the Milky Way \citep[e.g][]{winget1987, BSL, Liebert2005, Tremblay14}. 
As a by-product, we also identify cataclysmic variables (CVs) --- white dwarfs with ongoing mass exchange from a companion, and presenting emission lines, generally of hydrogen and/or helium --- and dC stars, due to their similarity with carbon-rich white dwarfs. These dC stars \citep{Roulston18}, as well as hot subdwarfs and ELMs,
hold potential to shed light on the poorly understood process of close binary evolution.

\section{Data analysis}

\subsection{Identification of the candidates}
We started with the 4\,851\,200 optical spectra in the SDSS DR14.
We selected the 259\,537 spectra of stars with $3\sigma$ proper motion larger than 20~mas/yr,
as well as all 68\,836 newly observed spectra of stars with colours within the \citet{dr7} selected white dwarf colour
range, and all 225\,471 spectra classified by the SDSS spectral pipeline as WHITE\_DWARF, A, B, OB or O stars, or
CV (cataclysmic variables). 
In addition, we performed an automated search for similar spectra as described in \citet{dr10,dr12} on all the 4\,851\,200 optical spectra, selecting further $\approx$ 4\,000 spectra.
We examined these selected spectra by eye ($\approx$ 500\,000 spectra, 
given the overlap between the different selections)
to identify broad line spectra characteristic of
white dwarfs, hot subdwarfs, and dCs, resulting in our identification of
34\,321 high signal-to-noise (S/N$_g$) spectra containing white dwarf, subdwarf, CVs and dCs stars. 
S/N$_g$ is the signal-to-noise parameter in the g-band in the SDSS spectra reduction
pipeline.
Our visual inspection showed that most objects
in the SDSS catalogue with proper motion smaller than 30~mas/yr
and magnitude $g>20$ are in fact galaxies, from their composite spectrum, high red-shifted lines, or broad emission lines.
We also inspected 1449 additional spectra for {\it Gaia} DR2 stars in the colour--magnitude white dwarf region [$M_\mathrm{GG}>3.333\times (\mathrm{G_{BP}-G_{RP}}) + 8.333$], not included in our previous selection. 
This white dwarf region was selected using the photometric conditions in \citet{Kilic18}, but with parallax/error > 4, flux/error > 3, as we are looking for stars with spectra and SDSS photometry, matching to 3 arcsec in the SDSS coordinates.

In previous SDSS white dwarf catalogues, we had not employed a proper motion criterion for selection, obtaining a signal-to-noise limited sample, determined by a colour---magnitude selection. The main reason we expanded our selection to include low signal-to-noise spectra from high proper motion
objects is that our previous colour
selection excluded the low temperature white dwarfs ($T_\mathrm{eff} < 8\,000$~K), because their SDSS colours are similar to the more numerous cool dwarf stars. 
However, considering that all stars born more than 2~Gyr ago with masses larger than $\sim 1.5~M_{\sun}$ are now white dwarfs cooler than 10\,000~K, our colour selection was excluding a significant population of these objects.
We still limited our classification to spectra with S/N$_g \geq$ 3--7, depending on the spectral type --- down to lower S/N$_g$ for DA stars because hydrogen lines are stronger and easier to detect, but to higher S/N$_g$ in other classes.

\subsection{Spectral Classification}

DR14 uses improved flux-calibration, with atmospheric differential refraction corrected on a per-exposure basis following the recipe described in \citet{Margala16},
and improved co-addition of individual exposures.
The Stellar Parameters Pipeline, which we used for our initial spectral class selection, are from \citet{lee08a, lee08b} and  \citet{AllendePrieto08}.

The wavelength coverage	is from 3800 to 9200~\AA\ for the SDSS spectrograph (up to Plate 3586), and 3650 to 10\,400~\AA, for the BOSS spectrograph, with a
resolution	of 1500 at 3800~\AA\ and 2500 at 9000~\AA, and a
wavelength calibration better than 5 km/s.
All the spectra used in our analysis
were processed with the spectroscopic reduction pipeline version
v5\_10\_0 for BOSSS/SEQUELS/eBOSS, the spectroscopic reduction pipeline
version 26 for the SDSS Legacy and SEGUE-1 programs, the special
SDSS pipeline version 103 to handle stellar cluster plates, and the
pipeline version 104 run on SEGUE-2 plates.
These
RUN2D numbers denote the version of extraction and redshift-finding code used.
In all SDSS spectral line descriptions, vacuum wavelengths are used.
The wavelengths are shifted such that measured velocities are relative to the solar system barycentre at the mid-point of each 15-minute exposure. 

Because we are interested in obtaining accurate mass distributions for
our DA and DB stars, we were conservative in labelling a spectrum as a clean
DA or DB,  adding additional subtypes
and uncertainty notations (:) if we saw signs of other elements, unresolved companions, or
magnetic fields (H) in the spectra.  While some of our mixed white dwarf
subtypes would probably be identified as clean DAs or DBs with better
signal-to-noise spectra, few of our identified clean DAs or DBs would
likely be found to have additional spectral features within our detection
limit.

We looked for the following features to aid in the
classification for each specified white dwarf subtype:

\begin{itemize}
\item Balmer lines --- normally broad and with a steep Balmer decrement
[DA but also DAB, DBA, DZA, and subdwarfs]
\item HeI $4\,471$\AA\ [DB, subdwarfs]
\item HeII $4\,686$\AA\ [DO, PG1159, sdO]
\item C2 Swan band or atomic CI lines [DQ]
\item CaII H \& K  [DZ, DAZ, DBZ]
\item CII $4\,367$\AA\ [HotDQ]
\item Zeeman splitting [magnetic white dwarfs]
\item featureless spectrum with significant proper motion [DC]
\item flux increasing in the red [binary, most probably M companion]
\item OI $6\,158, 7\,774, 8\,448$\AA\ [DS, oxygen dominated]
\item H and He emission lines [CVs and M dwarf companions]
\end{itemize}


Table~\ref{tab:all} is a tally of the 37\,053 objects we classified in Table~\ref{tab:all.1}. As 15\,716 
objects were classified by us as DAs and 1363 as DBs, of the 20\,109 white dwarfs in the table, 78\% are DAs. 

\begin{table}
	\centering
\caption{Classification of 37\,053 spectra in Table~\ref{tab:all.1}.
	\label{tab:all}}
	\begin{tabular}{ll} 
		\hline
Number & Type \\
		\hline
   15716 & DA \cr
    1358 & DB \cr
    1847 & DC \cr
     524 & DQ \cr
     598 & DZ \cr
      45 & DO/PG1159/O(He)/O(H) \cr
     210 & sdB \cr
     205 & sdO \cr
     311 & CV \cr
      4   & DS \\
      1   & DH \\
     14  & BHB \\
      & \cr
   15793 & sdA \cr
     447 & dC \cr
     8 & BL LAC \\
		\hline
	\end{tabular}
\end{table}

Among the 15\,716 DAs, we found 474 magnetic DAHs, 598 unresolved binaries with main-sequence M dwarf companions (DA+M), 136 DAZs with Ca and/or Mg lines, and 52 DABs contaminated by He~I lines.
We also found 41 stars having an extremely steep Balmer decrement (i.e. only a broad H$\alpha$ and sometimes H$\beta$ is observed while the other lines are absent)
that could not be fit with a pure hydrogen grid (see section~\ref{models} below), or indicated extremely high gravities.
We find that these objects are best explained as helium-rich DAs, and therefore with an extremely thin H layer mixed with the underlying He, and denote them DA(He).

\begin{figure}
   \centering
   \includegraphics[width=\linewidth]{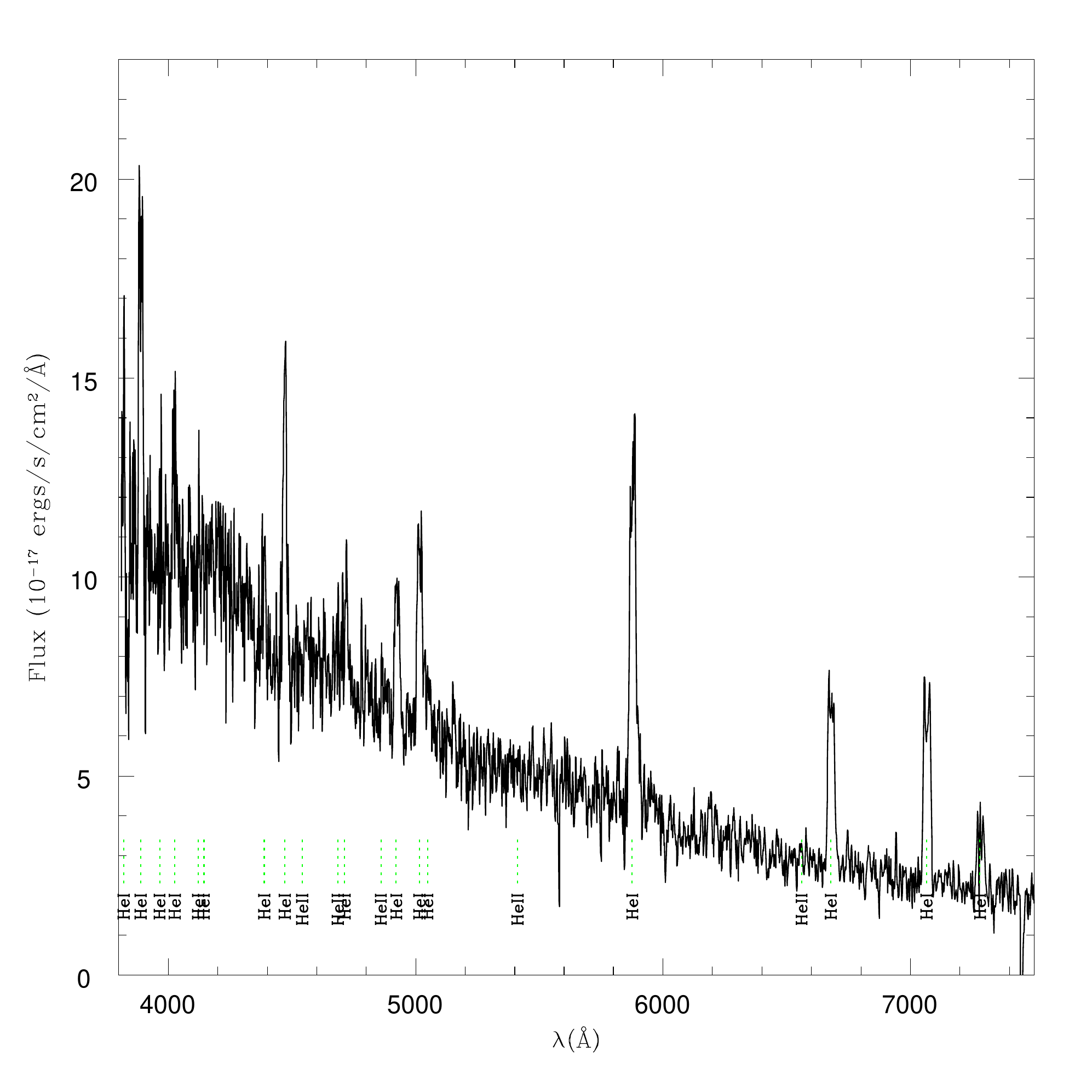}
      \caption{Spectrum of the outbursting AM CVn SDSS J141118.31+481257.66, with He emission double lines.}
      \label{Fig:amcvn}
         \end{figure}

We classified 447 spectra as dC - dwarf carbon stars, in line with \citet{Green13} and \citet{Farihi18}.
We cannot identify a clear visual discontinuity from the coolest DQs to the hottest dCs either in term of the C line strength or the colour--magnitude diagram (see Fig.~\ref{Fig:gaiahr}). We do not have spectral models for dCs, so we do not determine their properties.
Of the 340 CVs, 9 are AM CVn type, with pure He spectra, and 87 CVs show both H and He lines. As an example of the spectra of cataclysmic variables found in our search, Fig.~\ref{Fig:amcvn} shows the spectrum of the ultra-compact white dwarf binary AM CVn SDSS J141118.31+481257.66, with g=19.38, spectrum
P-M-F 1671-53446-0010, with He emission double lines (Fig.~\ref{Fig:amcvn}). 
\citet{Rivera18} reported an outburst,
the first recorded for this star. \citet{Ramsay18} review of AM CVns show many have outbursts reported;
AM CVn are ultra-compact hydrogen-deficient binaries, each consisting of a white dwarf accreting helium-dominated material from a degenerate or semi-degenerate donor star.

We classified 15\,855 stars as sdAs, stars with spectra dominated by narrow hydrogen lines, following \citet{dr12}. 
Solar metallicity main sequence A stars have absolute magnitudes $M_g\simeq 0$ -- 2. As stars brighter than g=14.5
saturate in SDSS, only A stars with distance moduli larger than 12.5 are observed in SDSS, i.e., farther than 3.5~kpc.   
Because SDSS observed mainly perpendicular to the disk (galactic latitude in general larger than $30\deg$), these would be located in the halo, where A stars should already have evolved off the main sequence.
Thus, these sdA stars are mostly likely very low metallicity main sequence stars ($\mathrm{[Fe/H]} \lesssim -1.0$), whose spectra are dominated by hydrogen because they lack significant metals, and most have masses smaller than the Sun, or their spectra
would show higher effective temperatures than observed. As their absolute magnitude, according to Gaia parallaxes, cover $-8 \geq M_\mathrm{G} \geq 10$ (see Fig.~\ref{Fig:sda}),
they cannot be classified as normal main
sequence A stars, presenting much lower masses and temperature than AV stars. They are hotter than sdF stars \citep{scholz15}. Some of these sdAs may be stars that lost mass due to binary interaction, resulting most probably in He core stars, precursors of ELMs, and ELMs
 \citep{Pelisoli18a,Pelisoli18b,Pelisoli18c} (see Section~\ref{section:sub}).
 \begin{landscape}
\begin{table}
\tiny
\begin{minipage}{\linewidth}
        \centering
        \begin{tabular}{llllllllllllllllll}
Plate-MJD-Fiber &      RA-DEC (2000)  & S/N$_g$ &  u     &$\sigma_u$&  g  &$\sigma_g$& r   &$\sigma_r$&  i  &$\sigma_i$& z  &$\sigma_z$&E(B-V)&ppm  & $\ell$&  b    & Type  \cr
                &                     &     & (mag)  &(mag)  & (mag)  & (mag) & (mag)  & (mag) & (mag)  & (mag)& (mag)  &(mag)  & (mag) &(mas/yr)&(deg) & (deg) &       \cr
2824-54452-0413&000003.29+282148.18&011&20.887&0.085&19.792&0.025&19.544&0.022&19.349&0.023&19.298&0.062&0.053&013.7&109.4&-33.2&sdA/F     \\
7850-56956-0719&000006.75-004653.98&022&19.236&0.034&18.855&0.024&18.914&0.016&19.123&0.021&19.447&0.049&0.038&002.1&095.7&-60.9&DA        \\
0650-52143-0497&000007.16-094339.84&007&19.429&0.043&19.577&0.028&19.997&0.025&20.275&0.040&21.212&0.415&0.029&040.5&085.7&-68.8&DA        \\
7134-56566-0587&000007.84+304606.35&011&20.214&0.039&19.666&0.017&19.531&0.021&19.499&0.023&19.567&0.059&0.039&044.7&110.1&-30.8&DA        \\
7666-57339-0848&000009.65+260022.32&004&21.119&0.034&21.104&0.045&21.104&0.045&21.231&0.076&20.994&0.233&0.036&000.0&108.8&-35.5&DC:       \\
8740-57367-0705&000010.29+064832.16&040&18.902&0.027&18.020&0.018&17.732&0.019&17.646&0.016&17.607&0.024&0.044&018.6&101.0&-53.9&sdA/F     \\
7167-56604-0752&000011.66-085008.31&019&19.453&0.039&19.111&0.021&19.065&0.023&19.139&0.021&19.338&0.083&0.032&104.7&087.0&-68.1&DQ        \\
4354-55810-0324&000012.04-030831.36&009&20.460&0.058&20.015&0.032&19.939&0.026&19.952&0.033&19.933&0.089&0.033&052.5&093.7&-63.1&DA        \\
7596-56945-0860&000012.57+190213.77&003&22.032&0.262&21.434&0.054&21.656&0.090&22.314&0.254&21.984&0.644&0.030&000.0&106.5&-42.2&sdA       \\
7167-56604-0246&000015.33-105859.15&067&16.616&0.020&15.685&0.026&15.353&0.021&15.242&0.017&15.184&0.017&0.030&032.9&083.8&-69.9&sdA       \\
0650-52143-0217&000022.54-105142.18&011&19.310&0.034&18.912&0.027&18.823&0.022&18.894&0.021&18.962&0.059&0.030&051.1&084.0&-69.8&DA        \\
8740-57367-0716&000022.68+064312.90&019&19.910&0.038&19.514&0.021&19.627&0.023&19.811&0.024&21.759&0.475&0.046&024.3&101.0&-54.0&DAH       \\
7848-56959-0062&000022.88-000635.69&049&18.270&0.019&18.247&0.027&18.571&0.016&18.858&0.027&19.222&0.055&0.029&016.2&096.4&-60.3&DA        \\
2822-54389-0400&000025.40+251726.40&010&20.243&0.045&20.194&0.027&20.502&0.026&20.825&0.046&20.986&0.190&0.042&012.4&108.6&-36.2&DA        \\
6127-56274-0026&000029.28+190638.88&007&23.157&0.045&19.780&0.028&19.780&0.028&19.524&0.038&19.509&0.114&0.031&000.0&106.6&-42.1&DC        \\
0650-52143-0165&000030.08-102420.70&029&18.374&0.021&17.481&0.022&17.215&0.013&17.101&0.019&17.111&0.019&0.034&022.2&084.9&-69.4&sdA/F     \\
2822-54389-0305&000030.24+242307.88&017&20.347&0.045&19.294&0.017&18.928&0.016&18.763&0.022&18.741&0.037&0.074&012.2&108.4&-37.0&sdA/F     \\
7850-56956-0688&000034.07-010819.97&037&18.189&0.015&17.844&0.021&18.013&0.023&18.240&0.020&18.502&0.035&0.032&030.3&095.7&-61.3&DA        \\
7034-56564-0336&000034.10-052922.46&046&16.930&0.023&16.778&0.020&17.060&0.016&17.296&0.015&17.576&0.021&0.029&086.6&091.4&-65.2&DA        \\
7850-56956-0704&000035.59-001115.88&007&23.039&0.347&20.271&0.031&18.813&0.017&18.334&0.026&18.118&0.039&0.030&065.8&096.5&-60.4&dC        \\
2824-54452-0272&000051.85+272405.26&025&18.702&0.029&18.648&0.022&19.063&0.023&19.304&0.029&19.618&0.096&0.045&013.4&109.4&-34.1&DA        \\
4354-55810-0735&000052.12-021437.58&008&20.810&0.081&20.191&0.023&19.944&0.023&19.993&0.032&19.772&0.081&0.031&080.4&094.8&-62.3&DC        \\
2803-54368-0210&000053.33+270330.00&067&16.723&0.015&15.647&0.015&15.618&0.015&15.611&0.017&15.714&0.026&0.037&026.2&109.3&-34.5&sdA       \\
0650-52143-0534&000054.38-090807.58&011&19.317&0.042&18.997&0.033&18.952&0.019&19.030&0.029&19.094&0.061&0.037&053.6&087.0&-68.4&DC        \\
7167-56604-0806&000054.40-090806.92&021&19.317&0.042&18.998&0.033&18.954&0.019&19.037&0.029&19.099&0.061&0.032&053.6&087.0&-68.4&DQ        \\
2630-54327-0342&000055.12-042449.04&010&19.771&0.043&20.016&0.032&20.113&0.026&20.373&0.043&20.642&0.163&0.041&029.3&092.8&-64.3&DBA:      \\
2630-54327-0359&000100.42-042742.87&028&18.814&0.029&18.513&0.043&18.707&0.020&18.882&0.026&19.165&0.047&0.038&016.9&092.8&-64.3&DA        \\
7848-56959-0026&000104.05+000355.82&035&19.220&0.028&18.867&0.028&19.051&0.018&19.212&0.027&19.482&0.053&0.025&010.5&096.9&-60.2&DA        \\
2822-54389-0397&000106.22+250330.05&017&19.516&0.035&19.526&0.023&19.711&0.022&19.970&0.033&20.246&0.135&0.063&015.8&108.7&-36.4&DBAZ      \\
2624-54380-0330&000106.77-034823.43&065&16.347&0.017&15.396&0.014&15.155&0.023&15.070&0.019&15.036&0.013&0.035&015.8&093.5&-63.8&sdA       \\
4534-55863-0466&000106.93+082825.58&019&19.401&0.028&18.946&0.017&18.848&0.015&18.840&0.018&18.922&0.046&0.049&067.1&102.3&-52.4&DA        \\
4216-55477-0238&000107.55-000042.83&021&19.924&0.033&19.033&0.028&18.695&0.022&18.599&0.021&18.591&0.039&0.027&013.9&096.8&-60.3&sdA/F     \\
1489-52991-0542&000108.80+001744.48&044&17.337&0.038&16.479&0.014&16.213&0.019&16.059&0.022&16.052&0.015&0.028&045.7&097.1&-60.0&sdA/F     \\
2824-54452-0207&000110.10+273520.40&011&20.335&0.038&20.071&0.025&20.071&0.025&20.139&0.040&20.461&0.161&0.043&009.8&109.5&-34.0&DA        \\
7850-56956-0267&000111.74-015620.55&031&19.049&0.023&18.208&0.022&17.870&0.014&17.765&0.015&17.736&0.022&0.031&017.2&095.3&-62.1&sdA/F     \\
2824-54452-0432&000115.77+285647.28&013&20.110&0.044&19.749&0.023&19.956&0.023&20.148&0.032&20.350&0.143&0.043&018.1&109.9&-32.7&DA        \\
6877-56544-0616&000115.78+261912.10&006&21.346&0.089&20.692&0.026&20.455&0.025&20.399&0.034&20.160&0.092&0.034&018.4&109.1&-35.2&DC:       \\

\end{tabular}
\caption{Spectral classification. The complete table is available electronically.
\label{tab:all.1}}
\end{minipage}
\end{table}
\end{landscape}

\subsection{Models\label{models}}

\begin{figure}
   \centering
   \includegraphics[width=\linewidth]{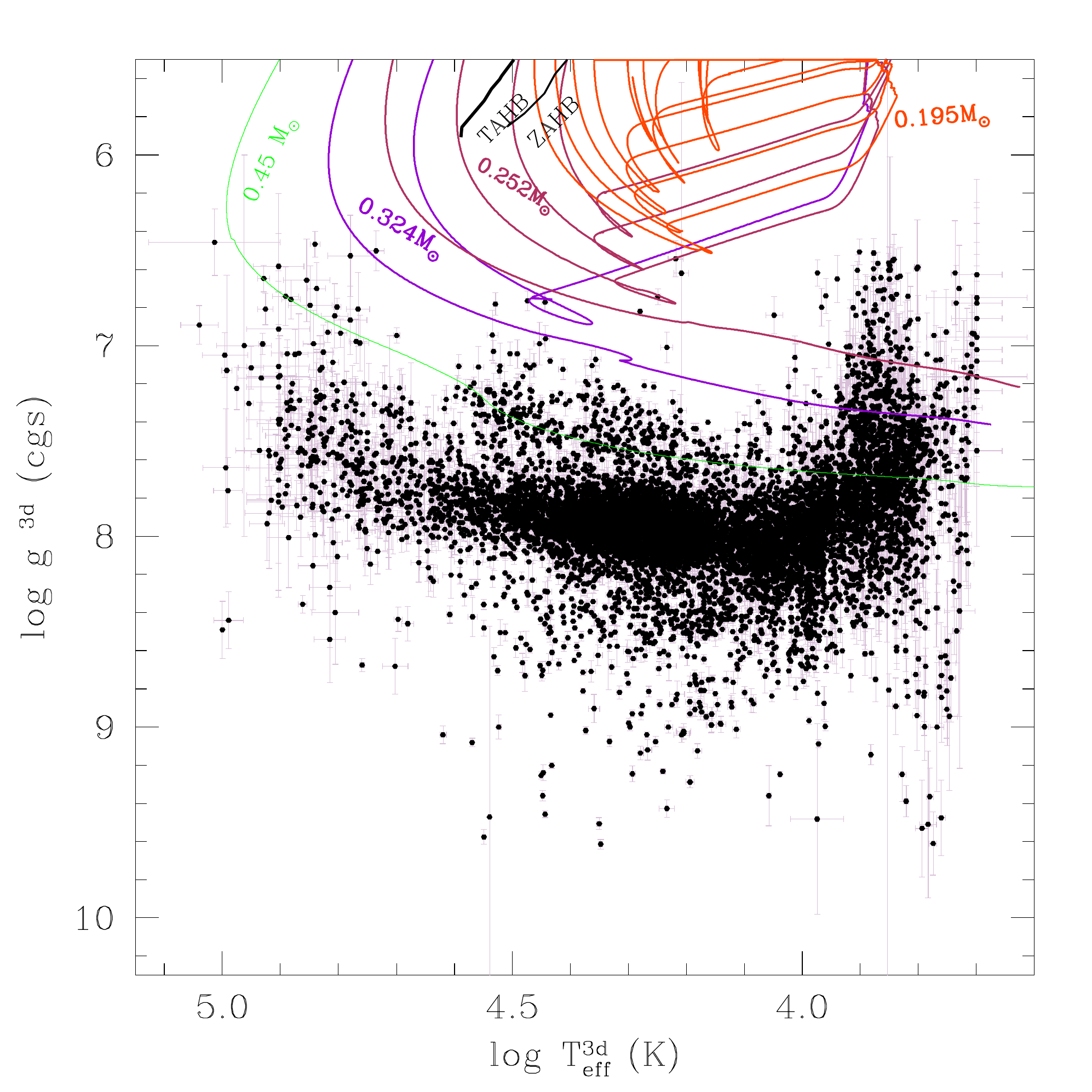}
      \caption{
Surface gravity ($\log g$) and effective temperature ($T_{\rm eff}$) estimated for the 10189 DA white dwarf stars for which the SDSS spectra has S/N$_g>10$, after applying three-dimensional convection atmospheric model corrections
from \citet{Tremblay13}, in black. 
The Zero Age Horizontal Branch (ZAHB) plotted was calculated with solar composition models. These
delimit the region of solar metallicity Blue Horizontal Branch stars.
It indicates the highest possible surface gravity for a hot subdwarf. Stars with
$T_\mathrm{eff}\leq 45\,000$~K and smaller surface gravities than the ZAHB are sdBs.
We have also plotted 0.45, 0.3, 0.2 and 0.15 $M_\odot$ models of He core pre-white dwarfs \citep{Althaus15,Istrate16} to guide the eye to the limiting region of what we call white dwarfs.
}
   \label{Fig:loggDA}
\end{figure}

After classifying white dwarf and subdwarf stars, we fitted the observed spectra to improved models of pure DAs and DBs \citep{Koester11,Koester15}, DOs \citep{Reindl14,Reindl14a,Reindl15}, sdBs and sdOs \citep{Geier15,Geier17,Geier17a}. 
For DAs, we used ML2/$\alpha=0.7$ models, with an LTE grid extending from 5\,000~K $\leq T_\mathrm{eff} \leq$  80\,000~K
and 3.5 $\leq \log g \leq$ 9.5 dex (cgs). For $T_\mathrm{eff} \leq 14\,000$~K we corrected the temperature and gravity to the 3D calculations of \citet{Tremblay13}, resulting in a flat $\log~g$ distribution down to $T_\mathrm{eff}\simeq 10\,000$~K, as shown in Fig.~\ref{Fig:loggDA}. The figure also show models for He core pre-white dwarfs \citep{Althaus15,Istrate16} and for the Zero Age Horizontal Branch (ZAHB) to show the region where we do not consider the objects as white dwarfs.
For DAs whose LTE analysis indicates $T_\mathrm{eff} \geq 45\,000$~K (that is where NLTE effects become important), we 
employed NLTE models. We computed a pure H grid with the T{\"u}bingen non-LTE Model-Atmosphere Package 
(TMAP, \citealt{werneretal2003, tmap2012, rauchdeetjen2003}) spanning from $T_\mathrm{eff}=40\,000-200\,000\,{\rm K}$ 
(step size 5000\,K for $T_\mathrm{eff} < 100\,000$\,K and $10\,000$\,K for $T_\mathrm{eff} > 100\,000$\,K) and $\log g=6.0-9.0$ 
(step size 0.5\,dex). To calculate synthetic line profiles, we used Stark line-broadening tables provided by \citet{TremblayBergeron2009}. 
To derive the effective temperatures and surface gravities the Balmer lines of the hot DAs were fitted in an automated procedure by 
means of $\chi^2$ minimisation using the FITSB2 routine \citep{Napiwotzki1999} and calculated the statistical one sigma errors. 
Each fit was then inspected visually to ensure the quality of the analysis. We excluded hot DAs whose spectra show an red excess and/or 
central emission features in the Balmer lines that cannot be the result of NLTE effects but are likely due the to the irradiation of a 
cool companion by the hot white dwarf.

\begin{figure}
   \centering
   
   \includegraphics[width=\linewidth]{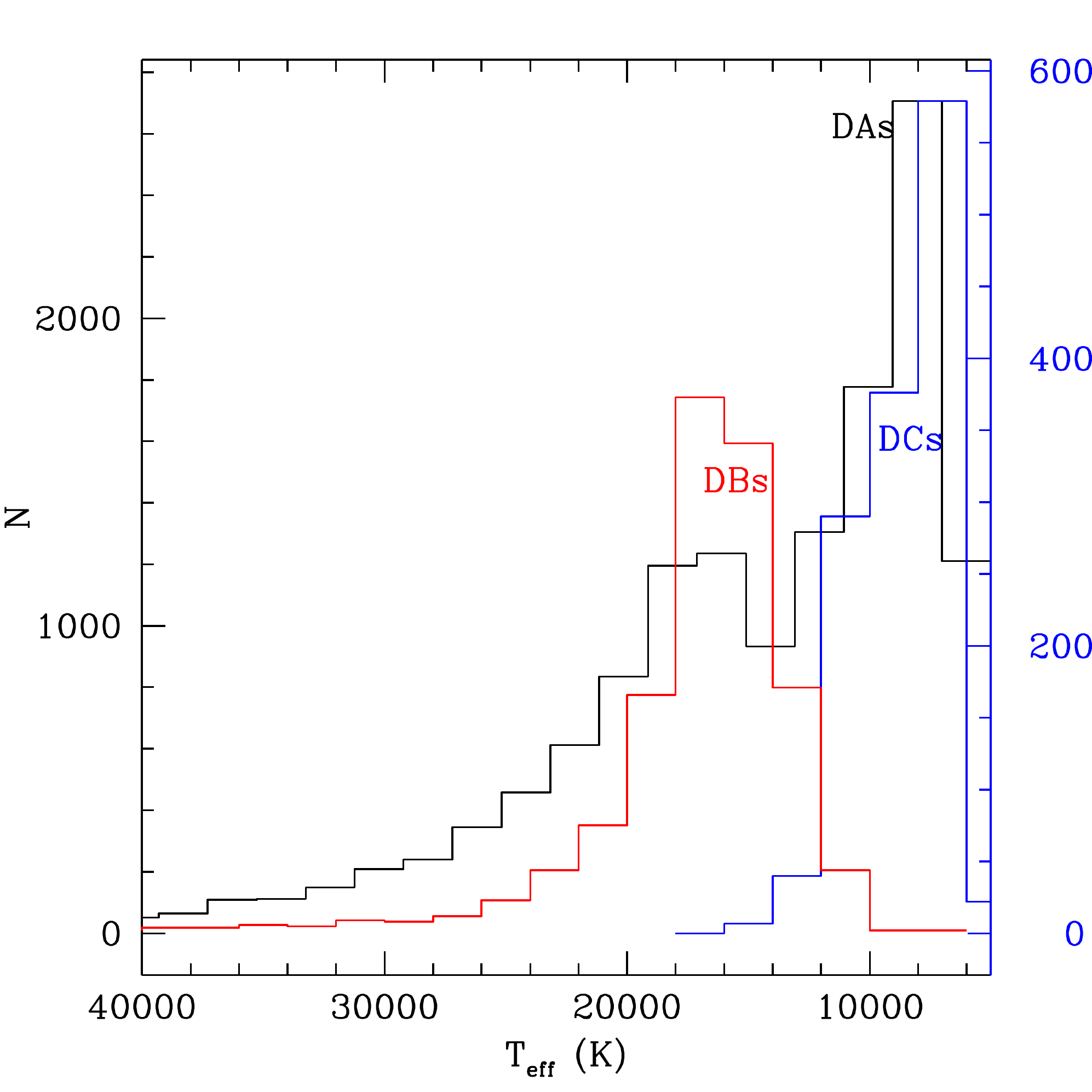}
      \caption{
Histogram of the number of DA stars versus effective temperature (in black), compared to the distribution for DCs (in blue), and DBs (in red). The number scale for DCs and DBs is shown on the right.
}
\label{histtdc}
\end{figure}

Fig.~\ref{histtdc} shows the histogram of the number of DA stars versus effective temperature. The hottest DAs we analysed have $T_\mathrm{eff}\simeq 120\,000$~K. The decrease of the number in the coolest bin is mainly due to incompleteness, because cooler stars are fainter --- partially compensated by the low mass stars that are brighter, but also affected by the finite age of the disk stars \citep{winget1987}.

\begin{figure}
   \centering
   \includegraphics[width=\linewidth]{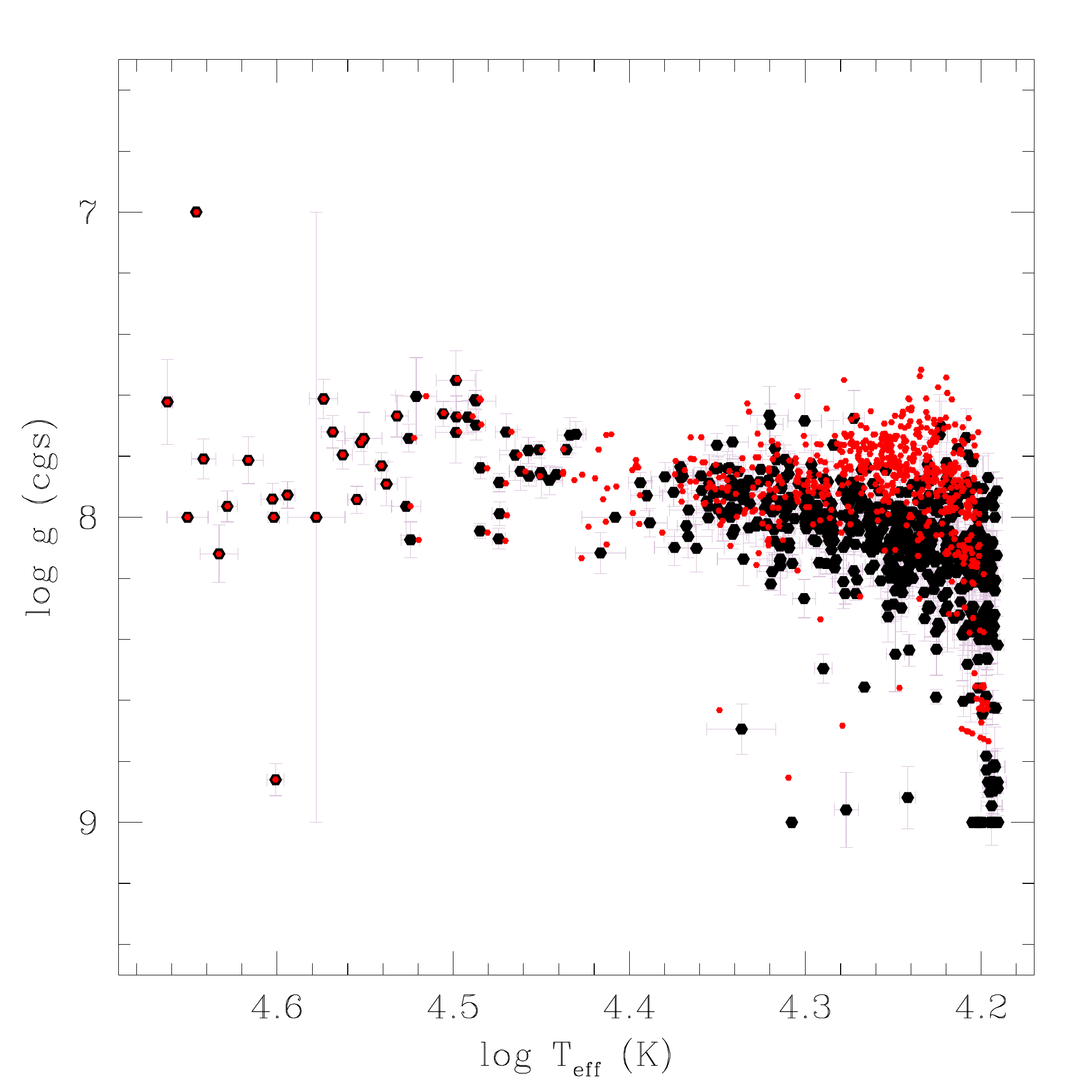}
      \caption{
Surface gravity ($\log g$) and effective temperature ($T_{\rm eff}$) estimated for 805 DB white dwarf stars with
spectral S/N$_g\geq 10$. The increase in apparent gravity below $T_\mathrm{eff}\simeq 16\,000$~K is probably caused by incorrect neutral broadening estimative \citep{Schaeuble17}. In red are the values after applying the pure He 3D corrections of \citet{Cukanovaite18}.
}
\label{dblogg}
\end{figure}

For DBs we use ML2/$\alpha=1.25$ LTE models as in \citet{Koester15}, with 12\,000~K $\leq T_\mathrm{eff} \leq$ 45\,000~K, and
7 $\leq \log g \leq$ 9.5 dex (cgs), resulting in the  $T_\mathrm{eff} - \log~g$ distribution shown in black in Fig.~\ref{dblogg}. An increase in the estimated $\log~g$ can be seen for $T_\mathrm{eff}\lesssim 16\,000$~K. This is not solved when pure He 3D corrections are applied \citep{Cukanovaite18}, shown in red in Fig.~\ref{dblogg}, and is probably caused by poor estimates of neutral broadening.

Because the spectral fits are normally degenerate between a hot solution(s) and a cool one, we also fitted the $ugriz$ colors of DAs and DBs to synthetic colours derived
from the same atmospheric models, and used the photometric values to guide our spectral parameter determinations.
For DCs, DQs and DZs, we only estimated their $T_\mathrm{eff}$ from the colours derived from the atmospheric models of \citet{Koester10}.
\section{Results}
\subsection{Masses\label{section:masses}}

For white dwarfs, the main indicator of $\log~g$ is the width of the atmospheric absorption lines. However, for $T_\mathrm{eff} < 10\,000$~K, the width of the hydrogen lines becomes very weakly dependent on gravity. As a result, it is very difficult to distinguish low mass white dwarfs and metal-poor main sequence A/F stars in the $T_\mathrm{eff} < 10\,000$~K and $\log g < 6.5$ range solely with visual inspection, even though low metallicity main sequence stars have an upper limit to $\log g\lesssim 4.64$, for a turn-off mass of $\sim 0.85~M_{\odot}$. 
The two steps we took to overcome this limitation was the extension of the model grid to $\log g\geq 3.5$, fitting all the spectra
we classified as DAs and sdAs, using the result to separate $\log g \geq 6.5$ as white dwarfs, and finally, after Gaia DR2, using the parallaxes, as discussed in Section~\ref{gaia}.

\citet{dr7} limited the white dwarf classification to surface gravity $\log g \ge 6.5$. 
At the cool end of our sample, $\log g=6.5$ corresponds to a mass around $0.2~M_\odot$, well below the single mass
evolution in the lifetime of the Universe --- but reachable via interacting binary evolution. The He-core white dwarf stars in the mass range $0.2-0.45~M_\odot$,
referred to as low-mass white dwarfs, are usually found in close binaries, often double degenerate systems \citep{Marsh1995}, being most likely a product of interacting binary stars evolution.
More than 70\% of those studied by \citet{Brown11} with masses below $0.45~M_\odot$ and all but a few with masses below $0.3~M_\odot$ show radial velocity variations \citep{Brown13, Gianninas14,Brown17}.
\citet{Kilic2007} suggest single low-mass white dwarfs result from the evolution of old metal-rich stars that truncate
evolution before the helium flash due to severe mass loss. They also conclude all white
dwarfs with masses below $\simeq 0.3~M_\odot$ must be a product of binary
star evolution involving interaction between the components.

The spectroscopic sdA sample defined in \cite{dr12} and used in our pre-selection here, being only a visual determination of narrow H lines and absence of strong metal lines, includes many types of objects: real white dwarfs, ELMs, pre-ELMs, low metallicity main sequence stars and even giants with low atmospheric metallicity. We need to define separate classes depending on the absolute luminosity (radius) to distinguish among them (see Section~\ref{section:sub}). Even though the nomenclature would suggest that subdwarfs have smaller radii than main sequence stars, it is not always the case --- they mainly have smaller masses.

\begin{figure}
   \centering
   \includegraphics[width=\linewidth]{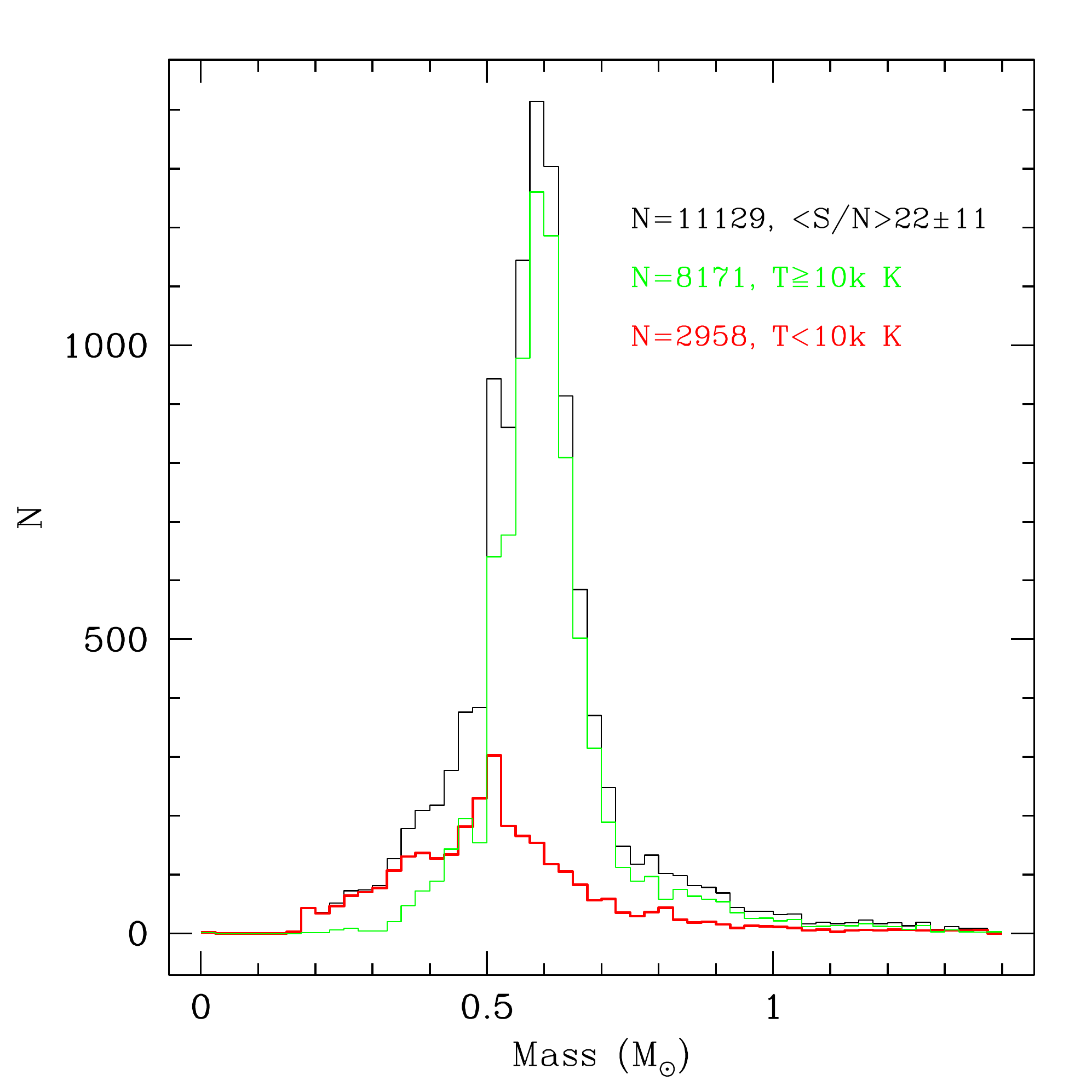}
      \caption{Mass distribution for the 11\,129 DAs with S/N$_g\geq 10$. This sample overall shows a mean mass of
      $\langle M \rangle=0.5904\pm 0.0014 M_\odot$. For stars with
      $T_\mathrm{eff} > 10\,000$~K, however, the mean mass is $\langle M \rangle=0.6131\pm 0.0014 M_\odot$, whereas for stars with  $T_\mathrm{eff} < 10\,000$~K, the mean mass is
      $\langle M_\mathrm{DB} \rangle=0.5276\pm 0.0035~M_\odot$. Most of the
      low mass DAs concentrate below $T_\mathrm{eff} = 10\,000$~K.
}
         \label{Fig:dahistm}
   \end{figure}

\begin{figure}
   \includegraphics[width=\linewidth]{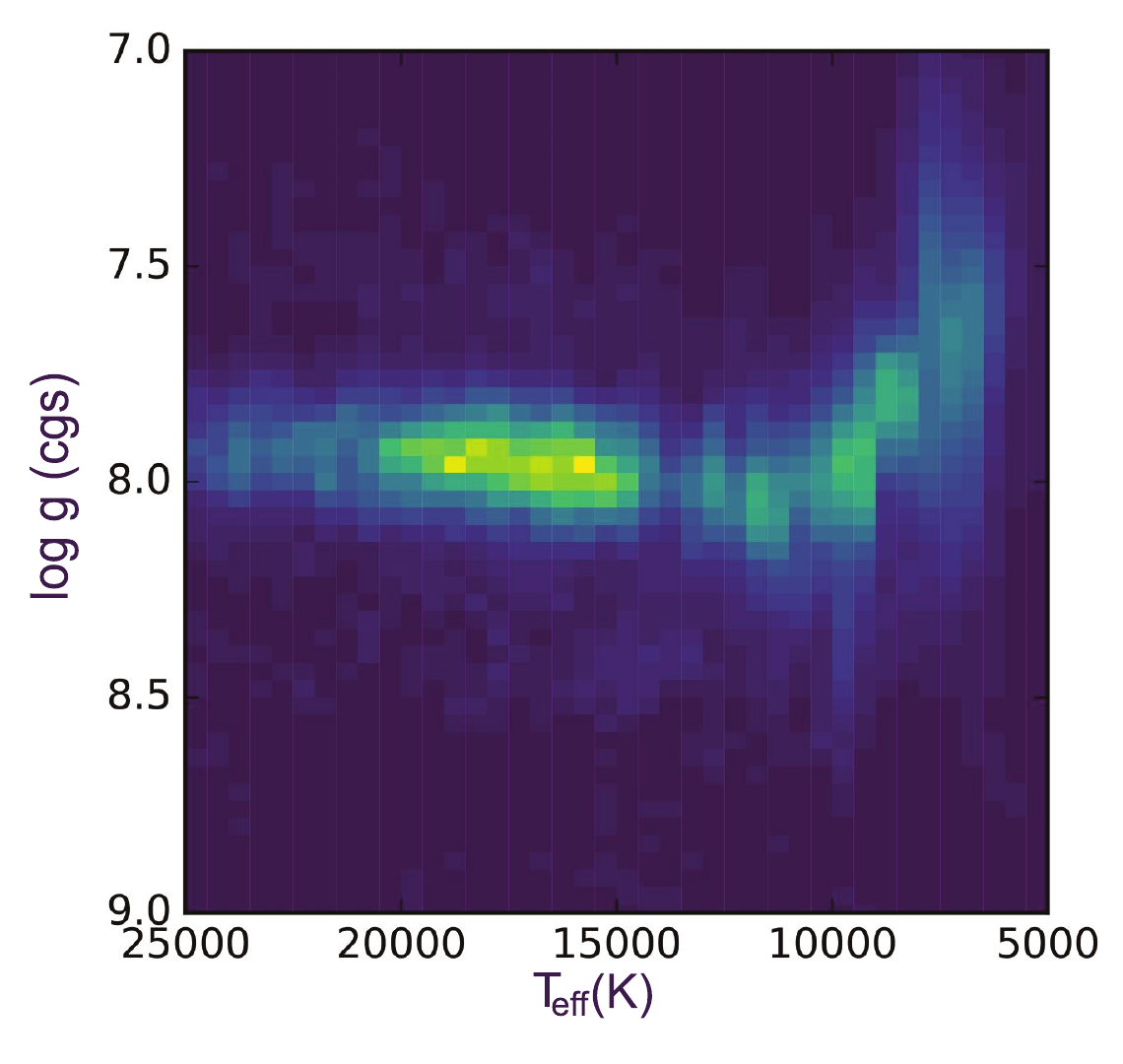}
      \caption{Hess diagram --- density distribution --- across effective temperature and surface gravity for DAs with spectra with S/N$\geq 10$, showing the
      low mass DA white dwarfs concentrate below $T_\mathrm{eff} = 10\,000$~K.
}
         \label{Fig:HessN}
\end{figure}

Table~\ref{tab:da} shows the atmospheric parameters obtained from the fitting of the spectra for DAs. Fig.~\ref{Fig:dahistm} shows the mass distribution for DAs with S/N$_g\geq$10, with 11\,129 stars, and result in a mean mass $\langle M_\mathrm{DA} \rangle=0.5903\pm 0.0014~M_\odot$, and individual dispersion of $0.152~M_\odot$. For the 8171 DAs with $T_\mathrm{eff}\geq 10\,000$~K, the mean mass is $\langle M_\mathrm{DA} \rangle=0.6131\pm 0.0014~M_\odot$, with a dispersion $0.126~M_\odot$, while for those 2958 with $T_\mathrm{eff}<10\,000$~K, $\langle M_\mathrm{DA}\rangle=0.5276\pm 0.0035~M_\odot$ with a dispersion $0.174~M_\odot$.
Fig.~\ref{Fig:HessN} shows the density of DAs versus temperature and surface gravity, showing the surface gravity decreases significantly below $T_\mathrm{eff}\simeq 10\,000$~K.
  
\begin{landscape}
\begin{table}
\tiny
\begin{minipage}{\linewidth}
        \centering
        \begin{tabular}{lllllllllllllllll}

Plate-MJD-Fiber&RA-DEC (2000) &Type&
$T_\mathrm{eff}$ (K)
&$\sigma(T)$&$\log g$&$\sigma(\log g)$&$V_r$ & $\sigma(V_r)$&d[pc]&z[pc]&
Mass$^\alpha=0.7$&
$\sigma(\mathrm{mass})$&
$T_\mathrm{eff}^{3D}$&
$\log g^{3D}$&Mass$^{3D}$&
$\sigma(\mathrm{mass}^{3D})$\\
0266-51602-0314&094107.47-001949.70&DA      & 011493&00122&8.396&0.079&  64&  22&00199&00118& 0.841&0.049&011437&8.235&0.73677&0.03484\\
0266-51630-0026&094901.28-001909.61&DA      & 010809&00027&8.094&0.025& -18&   5&00077&00048& 0.654&0.014&010702&7.875&0.53674&0.00847\\
0266-51630-0031&094804.30-000738.16&DA      & 010135&00050&8.525&0.068&  25&  14&00111&00069& 0.920&0.042&010033&8.255&0.74933&0.03043\\
0266-51630-0037&094917.06-000023.67&DA      & 011614&00152&8.437&0.095& -28&  23&00183&00114& 0.867&0.059&011568&8.281&0.77233&0.04207\\
0266-51630-0570&094640.35+011319.86&DA      & 019971&00126&7.922&0.022&  43&   7&00201&00125& 0.583&0.010&019971&7.922&0.58328&0.00723\\
0266-51630-0629&094844.84+003516.63&DA      & 012900&00106&8.146&0.034&  93&   9&00151&00094& 0.688&0.019&013007&8.105&0.66503&0.01347\\
0267-51608-0099&095329.20-005100.49&DA      & 011068&00113&8.653&0.092&  87&  23&00145&00091& 0.993&0.047&010952&8.438&0.86676&0.04082\\
0268-51633-0129&095759.16-010707.29&DA      & 007578&00028&7.708&0.076&  29&   5&00091&00058& 0.463&0.033&007574&7.567&0.40172&0.02207\\
0268-51633-0503&095940.23+003634.17&DA      & 010566&00069&8.049&0.078&  28&  15&00199&00130& 0.629&0.043&010452&7.800&0.51644&0.02100\\
0270-51909-0008&101435.26-004714.40&DA      & 009869&00073&8.497&0.120&   7&  22&00145&00099& 0.902&0.075&009783&8.232&0.73154&0.05228\\
0270-51909-0468&100955.38+000943.86&DA      & 010622&00035&8.170&0.037&  11&   7&00119&00081& 0.699&0.020&010512&7.922&0.56011&0.01355\\
0271-51883-0181&101741.70-002934.12&DA      & 014581&00094&7.993&0.018&   0&   0&00000&00000& 0.000&0.000&014581&7.993&0.60708&0.00657\\
0271-51883-0557&102003.39+000902.54&DA:     & 006148&00094&9.742&0.147&-188&  99&00000&00000& 1.372&0.034&006147&9.693&1.36091&0.02447\\
0272-51941-0289&102041.74-011101.13&DA      & 009211&00076&7.697&0.229&  55&  20&00214&00148& 0.464&0.093&009153&7.426&0.35929&0.05511\\
0272-51941-0307&101911.51+000017.25&DA      & 012777&00153&8.408&0.051& 129&  14&00156&00109& 0.850&0.032&012856&8.326&0.79890&0.02096\\
0272-51941-0518&102546.72+002857.43&DA      & 007890&00029&7.787&0.065&  21&   5&00092&00066& 0.500&0.027&007883&7.616&0.42446&0.01977\\
0273-51957-0337&102653.12+005110.54&DA      & 014768&00478&7.939&0.096&  84&  29&00322&00233& 0.580&0.048&014768&7.939&0.57950&0.03363\\
0273-51957-0615&103448.94+005201.33&DA      & 010070&00095&8.409&0.133& 170&  27&00166&00123& 0.847&0.082&009973&8.140&0.67917&0.05632\\
0274-51913-0222&103833.50-001959.70&DA      & 020825&00225&7.479&0.036&   0&   0&00000&00000& 0.000&0.000&020825&7.479&0.43104&0.00825\\
0274-51913-0303&103635.66-000036.42&DA      & 012283&00197&7.605&0.067&  27&  14&00335&00248& 0.440&0.026&012339&7.583&0.43140&0.01832\\
0274-51913-0536&104100.56+010909.44&DA      & 009563&00049&8.211&0.085&   9&  14&00168&00128& 0.718&0.054&009495&7.946&0.57061&0.03150\\
0276-51909-0073&105612.32-000621.66&DA      & 011167&00038&7.891&0.030&  32&   6&00149&00116& 0.546&0.015&011078&7.746&0.49279&0.00956\\
0276-51909-0097&105405.75-011132.96&DA      & 011621&00162&7.453&0.137&   0&   0&00000&00000& 0.000&0.000&011547&7.413&0.36752&0.02974\\
0276-51909-0593&105727.81+002118.70&DA      & 010731&00052&8.535&0.052&  52&  13&00145&00114& 0.927&0.032&010615&8.287&0.77429&0.02439\\
0277-51908-0025&110636.72-001122.37&DA      & 014717&00307&7.671&0.067&  17&  17&00305&00243& 0.475&0.025&014717&7.671&0.47527&0.01757\\
0277-51908-0066&110420.04-003628.21&DA      & 012528&00160&8.179&0.060&  41&  16&00203&00160& 0.707&0.035&012613&8.106&0.66488&0.02404\\
0277-51908-0414&110015.66+010740.55&DA      & 014348&00302&7.611&0.073&   0&  23&00407&00324& 0.452&0.027&014404&7.612&0.45284&0.01892\\
0277-51908-0513&110326.71+003725.80&DA      & 010636&00040&8.229&0.046&  -8&   9&00115&00091& 0.731&0.030&010527&7.981&0.59129&0.01728\\
0277-51908-0596&110515.32+001626.13&DA      & 013080&00049&8.275&0.011&  15&   3&00045&00036& 0.771&0.007&013180&8.235&0.74002&0.00577\\
0278-51900-0367&110623.40+011520.95&DA      & 010952&00066&7.943&0.063&   4&  13&00197&00159& 0.573&0.033&010848&7.759&0.49790&0.01833\\
0278-51900-0593&111230.14+003002.55&DA      & 009629&00038&8.159&0.062&  51&  10&00133&00108& 0.690&0.035&009556&7.894&0.54301&0.01928\\
0279-51984-0308&111028.70-003343.46&DA      & 009478&00048&8.299&0.091&   6&  14&00137&00109& 0.779&0.059&009416&8.033&0.61710&0.03553\\
0279-51984-0321&111047.52+005421.35&DA      & 008651&00026&8.127&0.052&  33&   6&00079&00064& 0.669&0.031&008628&7.877&0.53219&0.01434\\
0281-51614-0476&112712.17+001644.29&DA      & 012152&00269&8.116&0.119&  54&  33&00370&00308& 0.670&0.067&012200&8.026&0.61951&0.04541\\
0282-51658-0304&113036.48-002155.76&DAZ     & 005295&00201&6.765&0.537&   0&   0&00000&00000& 0.000&0.000&005294&6.761&0.18157&0.05746\\
0282-51658-0537&113614.89+005106.92&DA      & 009470&00062&8.564&0.095&  47&  19&00144&00123& 0.944&0.055&009409&8.299&0.77928&0.04347\\
0283-51584-0349&113901.22+000321.79&DA+M:   & 013256&00493&8.541&0.209&   0&   0&00000&00000& 0.000&0.000&013329&8.490&0.90135&0.08629\\
0283-51959-0117&114720.41-002405.66&DA      & 017517&00236&7.928&0.047&  70&  16&00318&00271& 0.581&0.023&017517&7.928&0.58051&0.01610\\
\end{tabular}
\caption{Parameters from spectral fitting to atmospheric models for DAs. The complete table is available electronically.}
\label{tab:da}
\end{minipage}
\end{table}
\end{landscape}

Table~\ref{tab:dat3d} shows the effective temperature distribution for DAs in our sample that were found to have effective temperatures in the range specified. The hottest we found is the DAO
SDSS~J160828.69+422101.77, with a S/N$_g=43$ spectrum and $T_\mathrm{eff}=120\,000 \pm 10\,000$~K,
while the hottest pure DA is SDSS~J101756.24+411524.72, with a S/N$_g=26$ spectrum and $T_\mathrm{eff}=110\,000 \pm 8\,000$~K. The most massive pure DAs are SDSS~J121234.85+165320.26, with a S/N=17 spectrum and $T_\mathrm{eff}=5\,944 \pm 91$~K, $\log g=9.611 \pm 0.166$, $M=1.370 \pm 0.006~M_\odot$, and SDSS~J152958.12+130454.80, with a S/N$_g=48$ spectrum and $T_\mathrm{eff}=5\,758 \pm 99$~K, $\log g=9.476 \pm 0.197$, $M=1.364 \pm 0.005~M_\odot$.
We caution that the quoted uncertainties are only the internal uncertainties from the least-square fits. For stars with multiple spectra, our mean external uncertainty is 5\% in the effective temperature and 0.05~dex in $\log g$, but for $T_\mathrm{eff} \leqslant 10\,000$~K, the real uncertainty is unknown. 

\begin{table}
        \centering
\caption{Distribution of DAs with $T_\mathrm{eff}$.}
        \label{tab:dat3d}
        \begin{tabular}{ll}
                \hline
   Number & Temperature Range \\
                \hline
    273   & < 6000 K\cr
   2938   & 6000 to 8000 K\cr
   2312   & 8000 to 10000 K\cr
   3325   & 10000 to 15000 K\cr
   3179   & 15000 to 20000 K\cr
   2968   & 20000 to 40000 K\cr
    596   & > 40000 K\cr
  15591   & total\cr
               \hline
        \end{tabular}
\end{table}
 
The histogram of the number of DB stars versus effective temperature can be seen in Fig.~\ref{histtdc} (in red), compared to for DAs (in black). We see no obvious DB gap, just the normal decrease in DBs hotter than 30\,000~K due to the ionization of HeI.  

    \begin{figure}
   \centering
   \includegraphics[width=\linewidth]{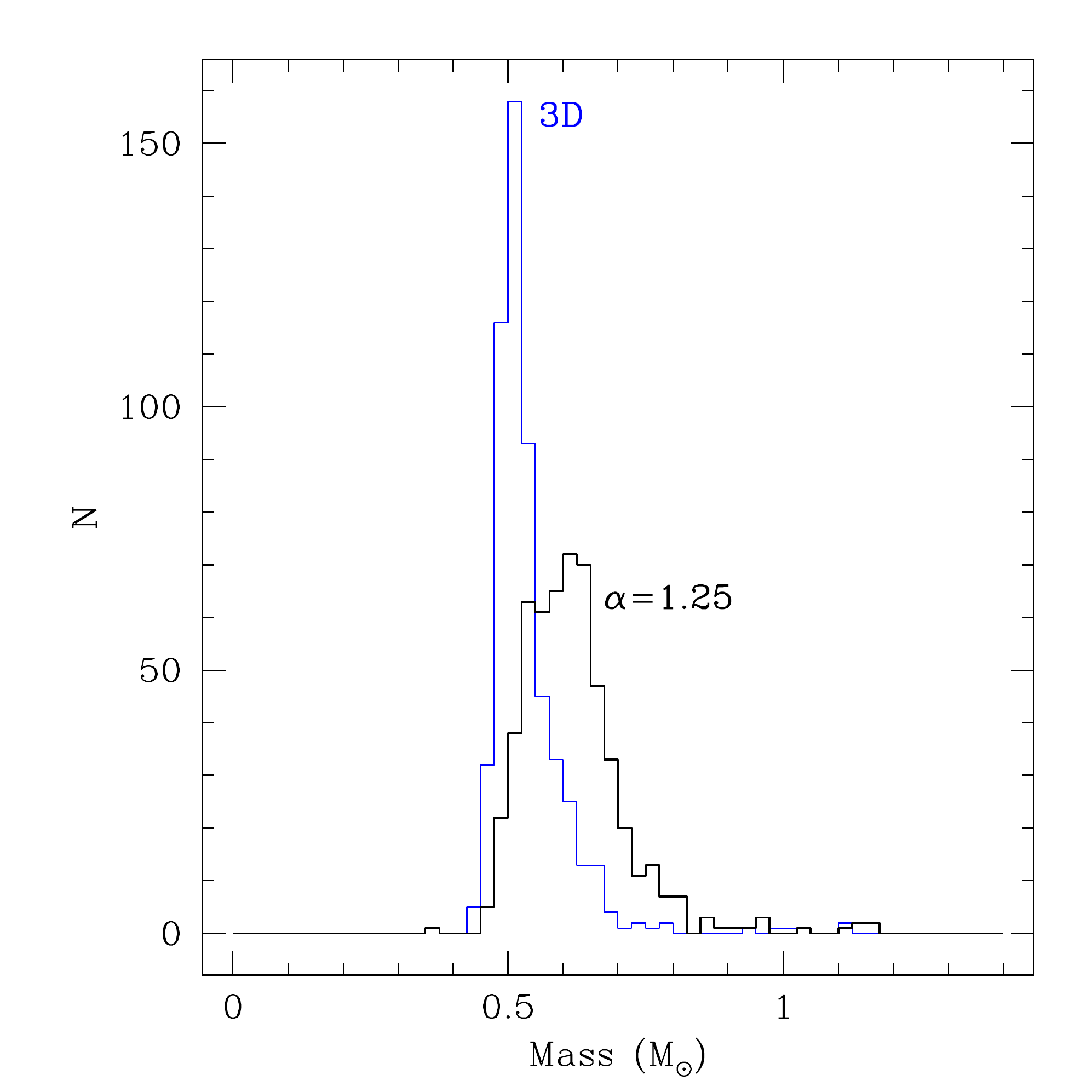}
      \caption{Mass distribution for the 550 pure DBs with S/N$_g\geq 10$ with $T_\mathrm{eff} > 16\,000$~K, the distribution shows a
      mean mass of $\langle M_\mathrm{DB}^{\alpha=1.25} \rangle=0.618\pm 0.004 M_\odot$, and a dispersion of $0.098~M_\odot$.
With the pure He 3D correction, in blue, the mean mass decreases to
       $\langle M_\mathrm{DB}^\mathrm{3D} \rangle=0.536\pm 0.003 M_\odot$, with a dispersion of $0.074~M_\odot$.
      For lower temperatures the $\log g$, and therefore mass, is not 
      trustworthy due to large uncertainties in the neutral broadening estimative. The DB mass distribution does not extend to masses below 0.45~$M_\odot$ or masses above 1.1~$M_\odot$; however, the statistics is much poorer than for DAs. The average signal-to-noise of the spectra is $\langle S/N_g \rangle = 25$. 
}
        \label{Fig:dbhistm}
         \end{figure}

Fig.~\ref{Fig:dbhistm} shows the mass distribution for the 550 pure DBs with S/N$_g \geq 10$ spectra and $T_\mathrm{eff} \geq 16\,000$~K, with and without the pure He 3D convection correction following \citet{Cukanovaite18}. Without the correction, the
mean mass is $\langle M_\mathrm{DB}^{\alpha=1.25} \rangle=0.618\pm 0.004 M_\odot$, and a dispersion of $0.098~M_\odot$.
With the 3D correction, the mean mass decreases to $\langle M_\mathrm{DB}^\mathrm{3D} \rangle=0.536\pm 0.003 M_\odot$, and a dispersion of $0.074~M_\odot$. The theoretical neutral broadening used in the models overestimates the $\log g$, and therefore masses, for lower temperatures \citep[e.g.][]{{Koester15,Schaeuble17}}. For the 333 pure DB with S/N$_g\geq 20$, we obtain $\langle M_\mathrm{DB}^\mathrm{3D} \rangle =0.533\pm 0.003~M_\odot$, with a dispersion of $0.058~M_\odot$,
i.e., the signal-to-noise is not changing the mean value. The low mean mass is a direct consequence of the pure He 3D corrections. Our fitted mean surface gravity, with the ML2/$\alpha=1.25$ models is
$\log g=8.032\pm 0.008$,
while the 3D corrected
$\log g=7.864\pm 0.007$.
A similar mean mass for DBs was obtained by \citet{genest19}.

The two highest mass DBs, above 16\,000~K, are
SDSS~J163757.58+190526.01, with S/N$_g=24$, $T_\mathrm{eff}=39895 \pm  441$~K, $\log g=8.86 \pm 0.05$, $M= 1.111 \pm 0.017~M_\odot$, and
SDSS~J081223.85+254842.82, with S/N$_g=14$, $T_\mathrm{eff}=20394  \pm 1000$~K, $\log g=8.85 \pm 0.05$, $M= 1.100 \pm 0.005~M_\odot$,
but our models, prior to the 3D correction, only go up to log g=9.0.

For the 1314 DCs in Table~\ref{tab:all.1} with Gaia DR2 parallaxes, we obtain their masses from the {\it ugriz} colours and Gaia DR2 parallax, 
following \citet{Ourique19}. Their radius is estimated from the observed flux and distance, assuming a He atmosphere mass-radius relation. Their distribution shows a mean surface gravity 
$\langle \log g ^{DC}\rangle=8.166 \pm 0.007$~dex~(cgs), with a dispersion of 0.245~dex,
and a mean mass of $\langle M^\mathrm{DC}\rangle=0.694 \pm 0.004~M_\odot$, with a dispersion of $0.127~M_\odot$.


   \begin{figure}
   \centering
   \includegraphics[width=\linewidth]{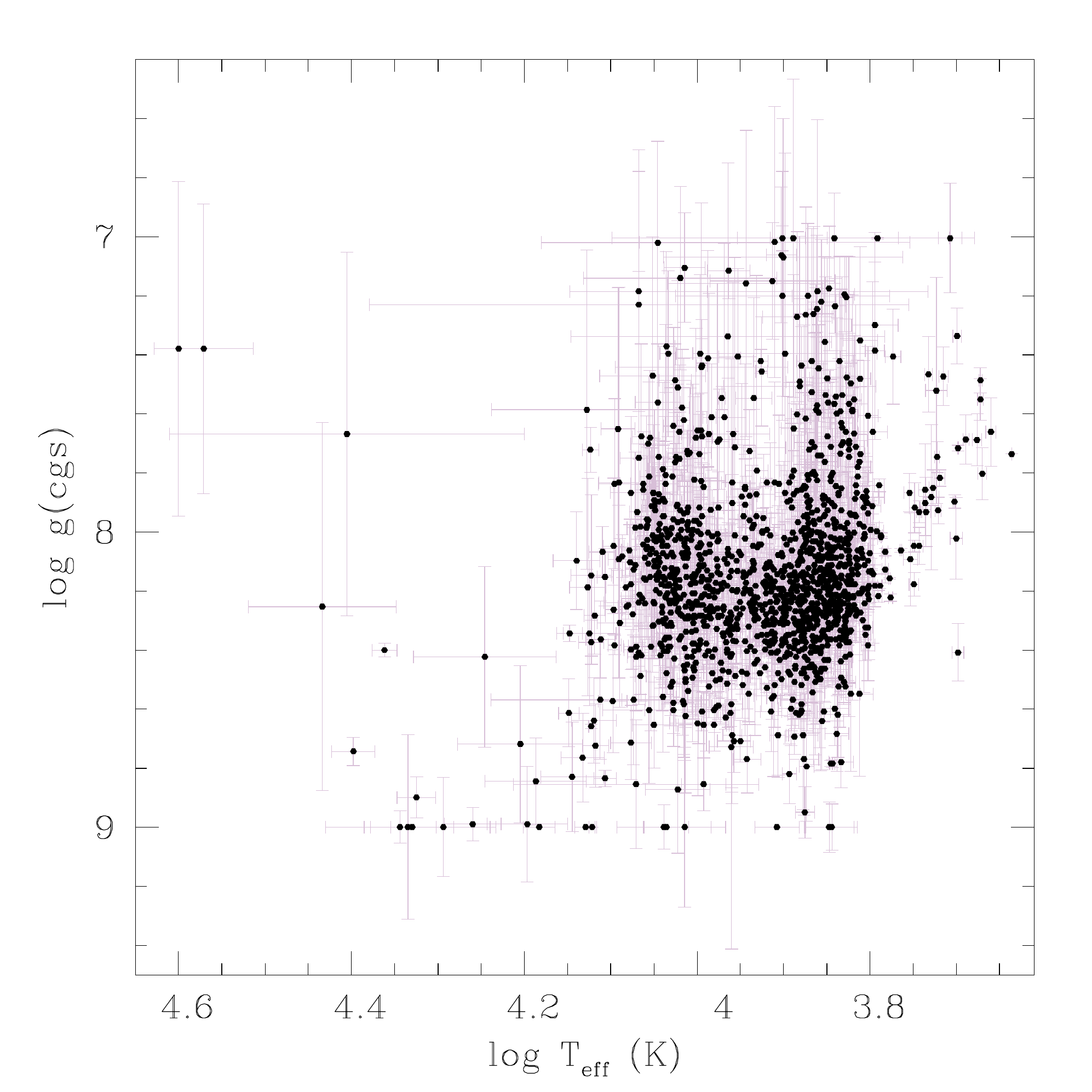}
      \caption{Surface gravity (log g) for the DCs, obtained from the SDSS ugriz photometry, Gaia DR2 parallax and a He-atmosphere mass--radius relation.}
      \label{Fig:dc}
         \end{figure}
         
Figure~\ref{Fig:dc} shows the effective temperature and surface gravity for 1314 DCs with Gaia DR2 parallaxes and {\it ugriz} {SDSS} colours. The mean mass obtained is $\langle M \rangle_\mathrm{DC}=  0.694 \pm 0.004~M_\odot$ with a dispersion of  $0.127~M_\odot$. \citet{Ourique19} presented the first DC mass distribution, using the Gaia colours and distances, showing it concentrates at higher masses than DBs. It is unlikely caused by an increase in the pressure by undetected hydrogen dredged up affecting the colours, as the non-DA to DA ratio increases below 16\,000~K.

\subsection{Hot White Dwarfs}
   \begin{figure}
   \centering
   \includegraphics[width=\linewidth]{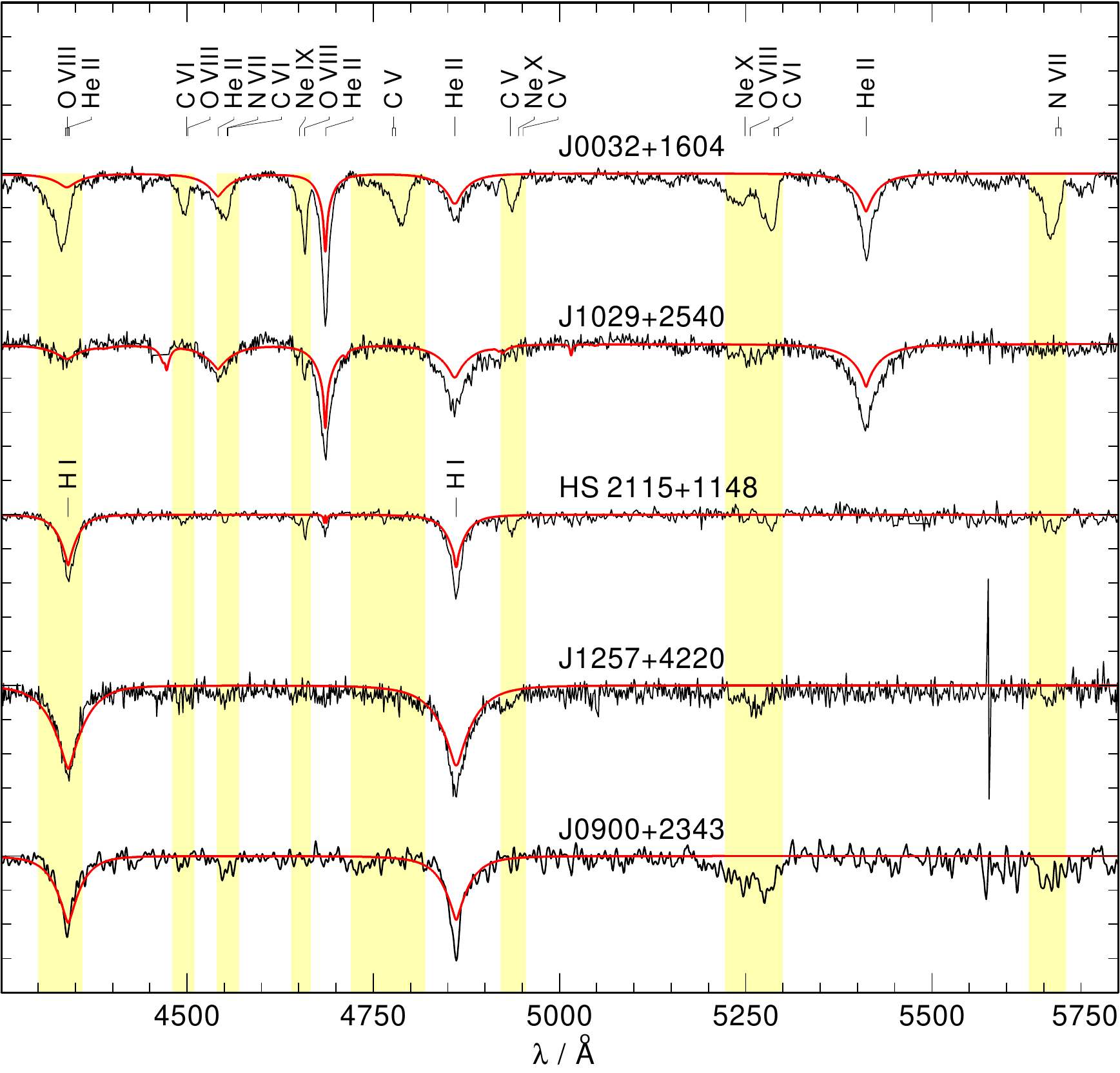}
      \caption{Normalized SDSS spectra of the newly discovered DO (upper two) and DA (bottom two) hot wind white dwarfs. 
      The spectrum of \mbox{SDSS\,J090023.89+234353.2} is convolved with a Gaussian (FWMH=3\AA) to smooth out the noise. The TWIN spectrum of HS\,2115+1148, for more than twenty years the only known H-rich hot wind white dwarf, is shown for comparison.}
      \label{Fig:hotwind}
         \end{figure}
We  identified spectroscopically in our sample a total of 12 PG~1159 and O(He) stars and 36 DOs with spectra 
dominated by He~II lines. Furthermore, we found one O(H) star \citep{Reindl16} and 48 DAO stars, with spectra showing both H and He~II lines as well as 310 hot DAs, 
showing only H lines. All these stars are hotter than $T_\mathrm{eff}=45\,000$~K, where NLTE effects are important in the spectral 
analysis. We note that the majority of these objects were already known and that our catalogue is far from being complete with 
respect to the hottest white dwarfs found in previous SDSS data releases. This is a consequence of the $3\sigma$ proper motion 
criterion, because hot white dwarfs are intrinsically more luminous and are detected over larger distances, and the more distant ones will have small proper motions.

One of the PG~1159 stars and ten of the DO white dwarfs belong to the group of the so-called hot-wind white 
dwarfs \citep{werneretal1995}, i.e. they show abnormally broad and deep He~II lines with seven of them showing additionally 
ultra-high excitation (uhe) absorption lines (e.g., O~VIII). For the latter objects we introduce the sub-classification uhe, i.e. 
PG~1159uhe and DOuhe. Our sample includes two newly identified hot wind DO white dwarfs, \mbox{SDSS\,J003213.13+160434.7}, which 
shows the strongest uhe features detected in any hot wind white dwarf so far, and \mbox{SDSS\,J102907.31+254008.3}, which shows 
only abnormally broad and deep He~II lines and possible an uhe feature located at 5250\AA. We also report the discovery of 
uhe features in two of the hot DA white dwarfs \mbox{SDSS\,J090023.89+234353.2} and \mbox{SDSS\,J125724.04+422054.2} 
(\mbox{PG 1255+426}). After more than twenty years, these are the first two H-rich hot wind white dwarfs discovered since
\mbox{HS\,2115+1148} \citep{Dreizleretal1995}. Figure~\ref{Fig:hotwind} shows the normalized SDSS spectra of the newly discovered objects. 
The uhe lines were recently shown to originate from an extremely hot, wind-fed circumstellar magnetosphere \citep{Reindletal2019}. 
The two newly discovered DA hot wind white dwarfs show the Balmer line problem \citep[failure to achieve a consistent fit to the Balmer lines,][]{Werner1996}, which is also present in \mbox{HS\,2115+1148}. Thus, the Balmer line problem can serve as a first indicator for the hot wind phenomenon. It is assumed that the cooler parts of the magnetosphere constitute an additional line forming region of the too-broad and too-deep H~I/-He~II lines \citep{Reindletal2019}.

\subsection{Subdwarfs}\label{section:sub}
We classified 77 stars as hot subdwarf sdOs, 128 sdOBs and 209 sdBs.
To refine the visual classification and derive the atmospheric parameters, a quantitative spectral analysis was performed for all sdO/B candidates in our sample with data of sufficient quality (S/N$_g$>20) and no atmospheric parameter determination in the literature. 

The method is described in \citet{Geier11}. We used appropriate model grids for the different sub-classes of sdBs and sdOBs. The hydrogen-rich and helium-poor $[\log{y}=\log{n(\rm He)/n(\rm H)}<-1.0]$ stars with effective temperatures below $30\,000\,{\rm K}$ were fitted using a of grid of metal line blanketed LTE atmospheres with solar metallicity \citep{Heber00}. Helium-poor stars with temperatures ranging from $30\,000\,{\rm K}$ to $40\,000\,{\rm K}$ were analysed using LTE models with enhanced metal line blanketing \citep{otoole06}. Metal-free NLTE models \citep{Stroeer07} were used for hydrogen-rich stars with temperatures below $40\,000\,{\rm K}$ showing moderate He-enrichment (log\,$y$\,=\,--1.0\,--\,0.0). The uncertainties provided are from statistical bootstrapping errors only. For more realistic uncertainties, additional random errors of about $\pm 1000\,{\rm K}$ in $T_{\rm eff}$ and $\pm 0.1\,{\rm dex}$ in $\log{g}$ should be adopted for sdBs and sdOBs. For the hotter sdOs $\pm2000\,{\rm K}$ and $\pm0.2\,{\rm dex}$ are more appropriate.

\begin{figure}
   \centering
   \includegraphics[width=\linewidth]{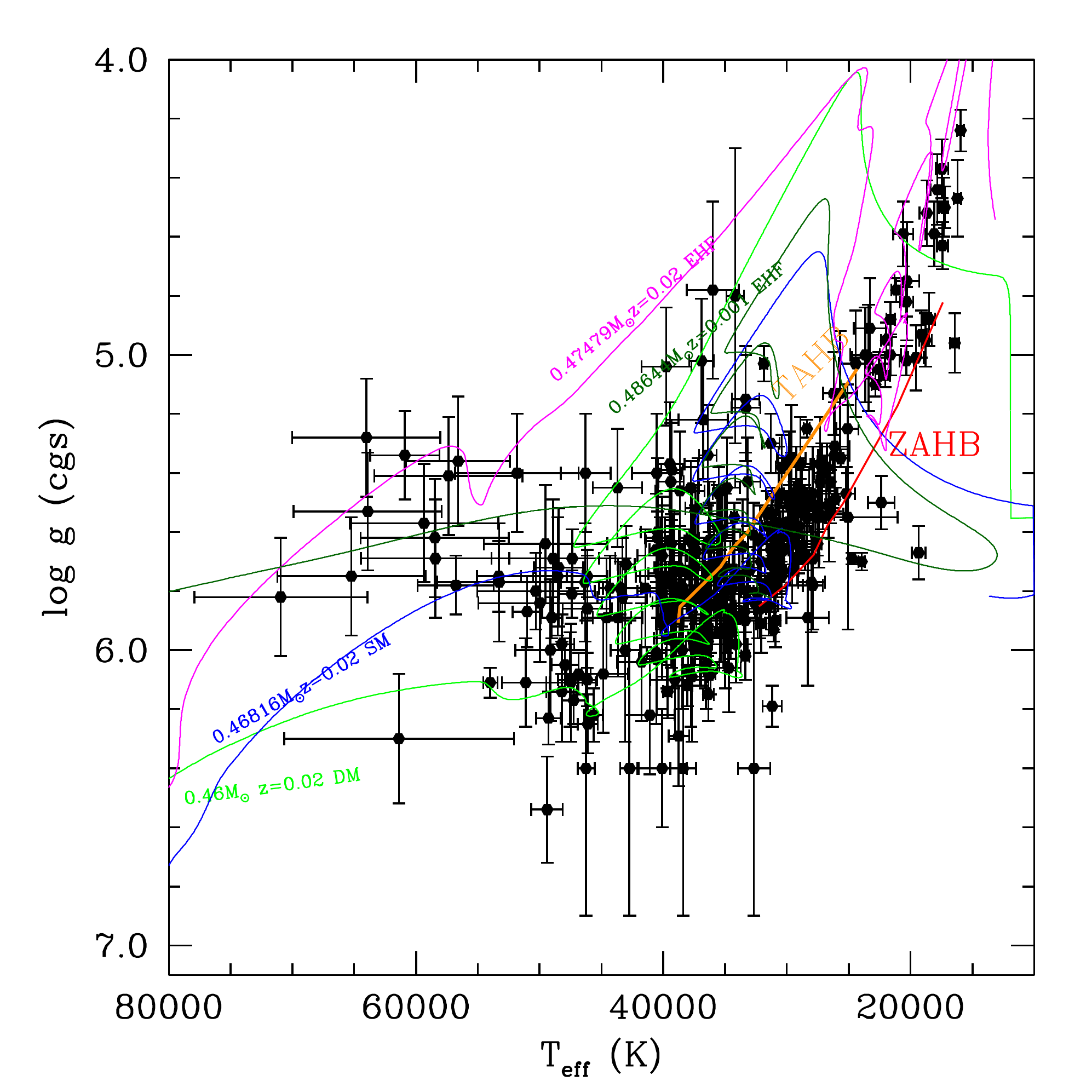}
      \caption{Hot subdwarfs, sdOs and sdBs. The Zero Age Horizontal Branch (ZAHB) and the Terminal Age Horizontal Branch (TAHB) plotted were calculated with solar composition models. In green we also plot a $0.46816~M_\odot$, z=0.02, SM (shallow mixing hot 
flasher),
DM (deep mixing hot flasher),
and EHF (early hot flasher) models from \citet{Tiara}.
      }
         \label{Fig:sdob}
   \end{figure}

Table~\ref{sdob} shows the hot subdwarfs analysed in this work, their classifications following the scheme proposed in \citet{Geier17} and their derived atmospheric parameters from the literature or this work. Fig.~\ref{Fig:sdob} shows effective temperature and surface gravity for the sample of hot subdwarf O- and B-type stars.

\begin{landscape}
\begin{table}
\caption{Table of Hot Subdwarfs}
\label{sdob}
\begin{minipage}{\linewidth}
        \centering
\begin{tabular}{llccccccccccccccccccccccccccccc}
\hline
\hline
P-M-F&SDSS J&Type&Teff&$e_\mathrm{Teff}^\mathrm{fit}$&$e_\mathrm{Teff}^\mathrm{total}$&
$\log g$&$e_\mathrm{logg}^\mathrm{fit}$&$e_\mathrm{logg}^\mathrm{total}$
&log n(He)/n(H)&$e_\mathrm{logn(He)/n(H)}^\mathrm{fit}$&$e_\mathrm{log n(He)/n(H)}^\mathrm{total}$&New/Known\cr
                \hline
4017-55329-0110&  151250.01-015436.33&BHB     &17321    &372       & 623        &4.50       &0.10      &0.14       &-2.46        &0.28               &0.34                  &New\cr
5325-55980-0738&  095638.14+145258.60&BHB     &17411    &449       & 672        &4.63       &0.08      &0.13       &-1.74        &0.18               &0.27                  &New\cr
5420-56009-0298&  130625.91+133349.14&BHB     &17247    &213       & 543        &4.50       &0.07      &0.12       &-2.23        &0.33               &0.39                  &New\cr
4775-55708-0626&  151519.21+054333.33&BHB     &15960    &274        &570        &4.24       &0.07      &0.12       &-2.01        &0.34               &0.39                  &New\cr
0793-52370-0623&  151847.69+551154.24&BHB     &18669    &604        &784        &4.52       &0.11      &0.15       &-1.99        &0.21               &0.29                  &New\cr
4504-55571-0996&  082216.14+133822.54&He-sdB  &32648    &1296       &1523       &6.40      &            &           &1.25         &0.09     &          0.22                  &New\cr
5064-55864-0668&  220711.11+125755.59&He-sdO  &53992    &565        &2078       &6.11       &0.05      &0.16       &2.00          &         &                               &New\cr
4493-55585-0560&  080833.77+180221.83&He-sdO  &46857    &966        &2221       &6.08       &0.07      &0.17       &2.00          &          &                              &New\cr
5200-56091-0132&  161023.39+371315.74&He-sdO  &47257    &958        &2218       &6.17       &0.08      &0.17       &2.00           &         &                              &New\cr
5172-56071-0644&  144321.34+402834.06&He-sdO  &49283    &987        &2230       &6.23       &0.09      &0.17       &2.00           &         &                              &New\cr
\end{tabular}
\end{minipage}
\end{table}
\end{landscape}

\begin{figure*}
   \centering
   \includegraphics[width=\linewidth]{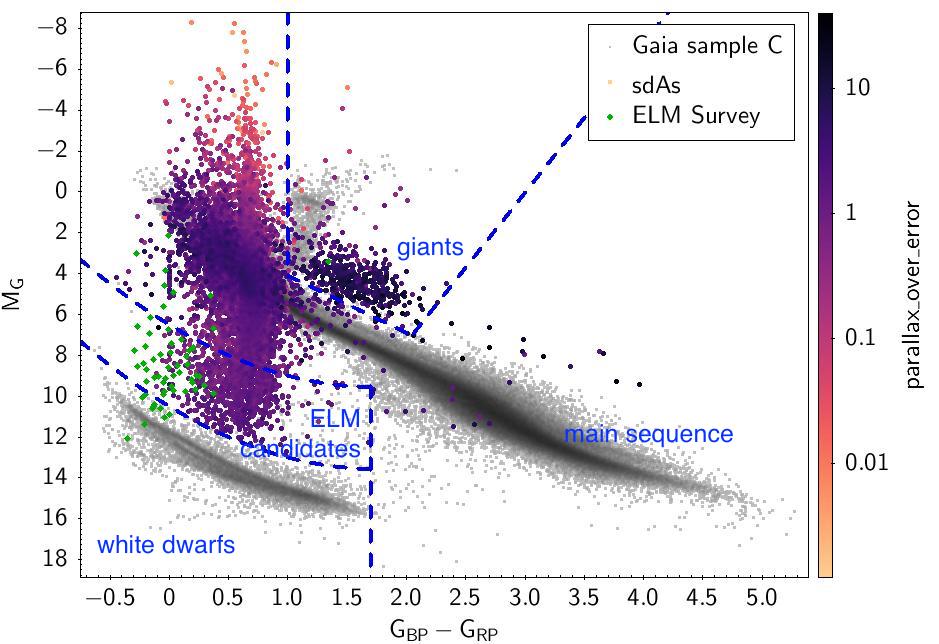}
      \caption{sdAs (colour-coded according to parallax\_over\_error) separated using their {\it Gaia} absolute magnitude and colour. They extend from the white dwarf cooling region, binary region, through the low metallicity main sequence, to the giant phase. As a comparison, we also show the sample C of \citet{lindegren2018}, consisting of solar neighbourhood stars (within 100~pc) with clean parallax, and the ELMs of \citet{Brown16}}.
      \label{Fig:sda}
      \end{figure*}

We also classified 15\,793 as sdAs, which is only a spectroscopic class to flag objects with narrow H lines \citep{dr12}. The classification carries no information on their origin or radius. The {\it Gaia} DR2 parallax determinations (next section) are somewhat uncertain for these objects (see Fig.~\ref{Fig:sda}), being in many cases of the same order of the error. This leads to a large scatter, placing objects above and below the main sequence, the latter a region compatible with low metallicity main sequence stars, or interacting binary remnants \citep[e.g.][]{Maxted14a,Pelisoli18a,Pelisoli18b,Pelisoli18c,vanRoestel18, Wang18}.
Some sdAs could be very low mass white dwarfs or pre-ELMs, but with the current parallax uncertainties this cannot be confirmed. We verified that by simply adding twice the parallax uncertainty to its value, over 95 per cent of the objects in the region between the main sequence and the white dwarf cooling track become compatible with the main sequence, suggesting that an inaccuracy of only 2-$\sigma$ in the parallax values is sufficient to explain this spread. Moreover, most of these objects have tangential velocities larger than 200 km/s, the criterion used for halo stars in the Gaia papers, providing further indication that they are compatible with low-mass main sequence stars in the halo. However, we caution that a high tangential velocity could also be observed for an object in a close binary, which is the case for the ELMs. In the next {\it Gaia} data releases, when the astrometry of binary objects and nearby contamination is more accurate, we will have a better understanding of the origin of these sdAs.

Fig.~\ref{Fig:sda} shows the $M_G$ vs. $G_\mathrm{BP}-G_\mathrm{RP}$ diagram of the identified sdAs (colour coded by parallax\_over\_error), with a tentative colour separation for canonical white dwarfs, (pre-)ELM candidates, and stars in the main sequence region and giant, which might be low-metallicity main sequence stars or binaries \citep[e.g.][]{Istrate16}.

\section{Gaia}
\label{gaia}
\begin{figure}
   \centering
   \includegraphics[width=\linewidth]{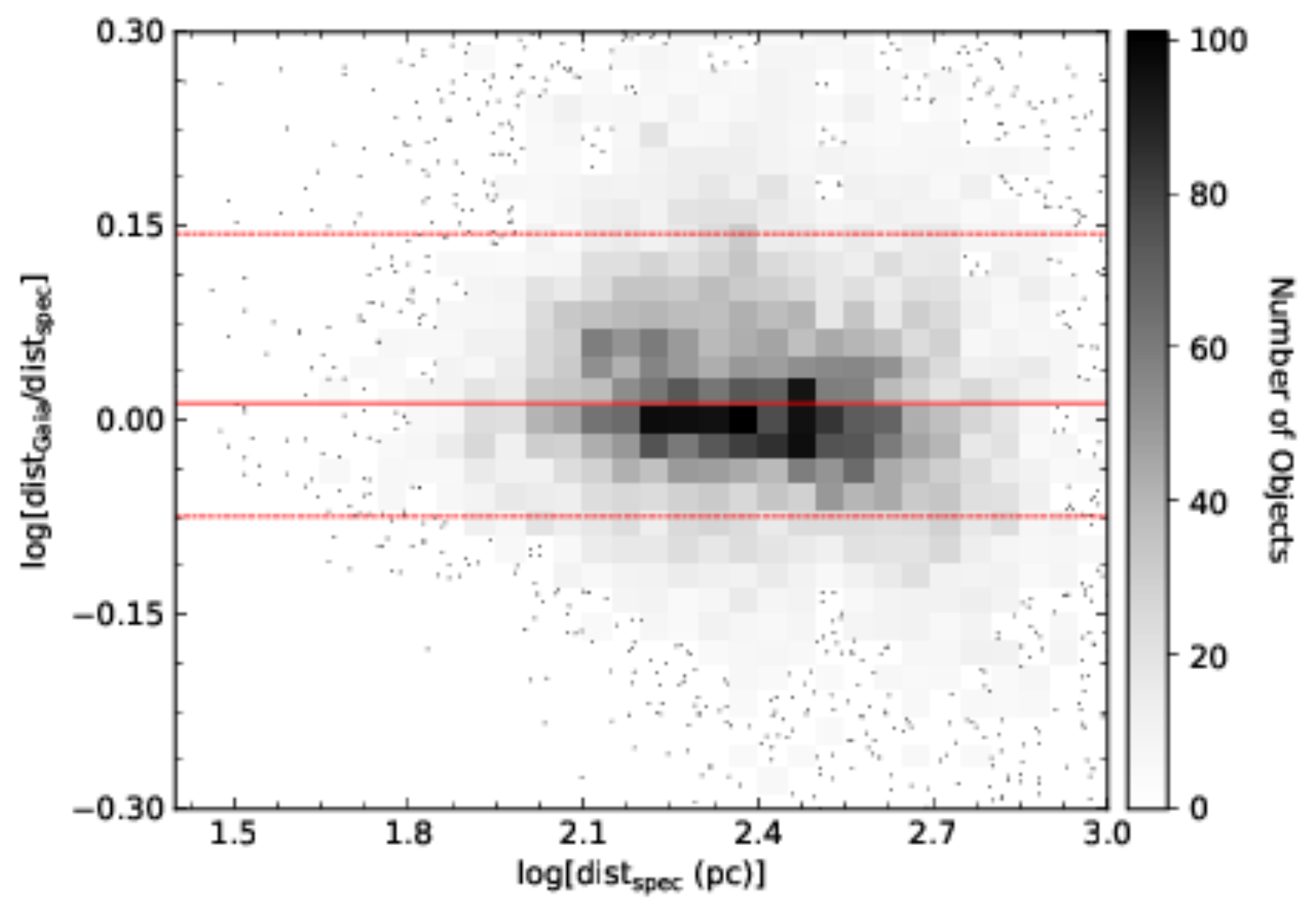}
      \caption{Distances for DA white dwarfs in our sample, estimated from the Gaia parallax uncertainty distribution, compared with the distance calculated from the spectroscopic distance modulus. The solid red line represent the median. The lower and upper dashed red lines represent, respectively, the 16 and 84 percentiles. The points represents bins with less than 5 objects. We did not use a Gaia DR2 parallax precision limit in this plot, or used the parallax to select the best spectroscopic solution. The distances from the parallax are compatible with the spectroscopic distances we obtained given the large error bars.}
         \label{Fig:gaiadist}
   \end{figure}

{\it Gaia} DR2 listed proper motion for 34\,499 of our objects, but did not obtain parallax for 4\,539 of these.
The proper motions were mainly compatible with those from the USNO, APOP and GPS1, and the distances from the parallax are compatible with the spectroscopic distances
we obtained.
Fig.~\ref{Fig:gaiadist} shows a comparison of the distances estimated from {\it Gaia} parallax by \citet{Bailer18} versus the distance estimated from our spectroscopic fits for DA stars, 
showing they are compatible but with a large scatter. The scatter is sometimes caused by the degeneracy of hot and cold solutions in the spectroscopic determination, and low S/N$_g$, but mainly above magnitude g=20 or distances larger than 1.5~kpc.

\begin{figure*}
   \centering
   \includegraphics[width=\linewidth]{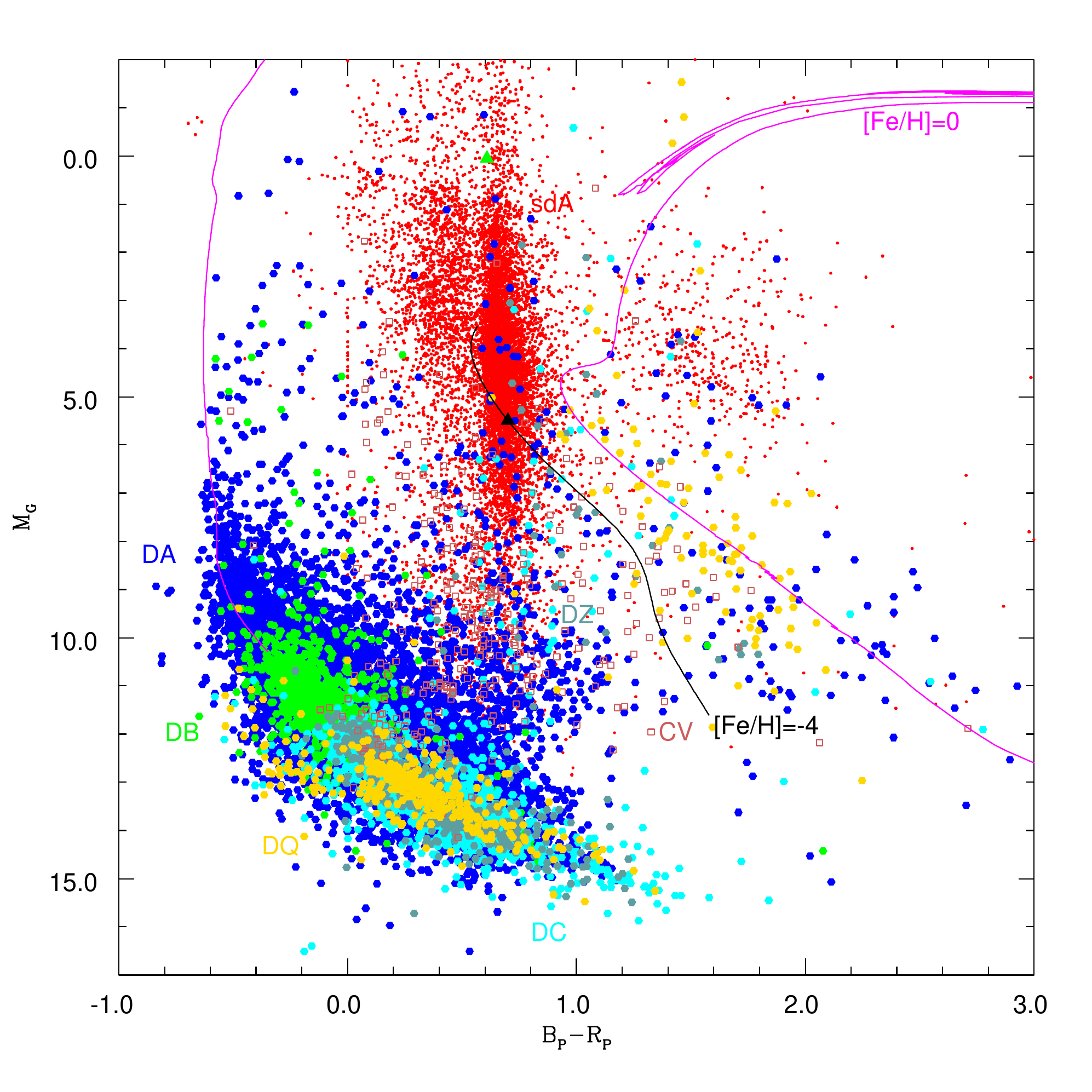}
      \caption{Colour $\times$ magnitude diagram of our sample, from Gaia distances and colours. The lines are 10~Gyr MIST isochrones with solar metallicity and [Fe/H]=-4 \citep{Choi18}. We also included two triangles, the upper one (green) with $T_\mathrm{eff}=6\,000$~K, 10~Gyr, [Fe/H]=-2.5, off the main sequence, and the lower one (black), for [Fe/H]=-4, still on the main sequence, for reference, from the MIST models.
      }
         \label{Fig:gaiahr}
   \end{figure*}

Fig.~\ref{Fig:gaiahr} shows the Hertzprung-Russell colour-magnitude diagram of our DR14 sample, using only the Gaia measurements, 
totally independent of our spectroscopic measurements. They
show DAs and DBs spread through the diagram, compatible with the \citet{Kilic18} conclusion that the gap seen in \citet{Babusiaux18} white dwarf HR diagram is not mainly due to atmospheric composition. \citet{elbadry18,elbadry18a,elbadry18b} used the main sequence --- white dwarf wide binaries with parallax/error $> 20$ for parallax$> 10$ mas ($d<100$~pc) and $M_G < 14$, corresponding to $T_\mathrm{eff} > 6000$~K, in {\it Gaia} DR2, to study the IFMR of white dwarfs, specially for initial masses $< 4M_\odot$, and conclude the bi-modality seen in the {\it Gaia} data constrains the data to multiple populations.

\section{Discussion}

The systematic uncertainties in our atmospheric parameters derived from spectral analysis are minimized by the use of only SDSS spectra, i.e., same telescope and only two
spectrographs (SDSS and BOSS), and fitting all the spectra with the same models and fitting technique.

\citet{nicola18} selected photometrically white dwarf candidates from the Gaia DR2 and classified those they matched to SDSS spectra in DR14, similar but a subset of our work. We matched their catalogue and we did not miss any star they classified, but we do include other objects they did not classify, because of their selection criteria.

\citet{Rolland18} analysed 115 helium-line DBs and 28 cool He-rich hydrogen-line DAs through $S/N\geq 50$ spectra and concluded 63\% of the DBs show hydrogen lines.
\citet{Koester15}, 
using S/N$_g\simeq 20$ SDSS spectra, measured 75\% DBs show hydrogen and speculated
all DBs show hydrogen, if observed at high resolution and S/N.
The surface gravity obtained with fits of pure He, for DBs containing H, and of pure H for DAs containing He, are overestimated, because of the extra particle pressure. \citet{Rolland18} conclude only if
$M_H\geqslant 10^{-6} M_*$ H will not mix with the underlying He layer by convective mixing at low effective temperatures. 
In our mean mass determination, we only included DBs hotter than $T_\mathrm{eff} \geq 16\,000$~K and that we did not see any contamination, but it is limited by our signal-to-noise and resolution.
\citet{Tremblay18} fitted 3171 $S/N\geq 20$ DR14 spectra for DAs and 405 DBs for the white dwarfs selected by \citet{nicola18},
applying 3D corrections for both DAs and DBs, as we did, and compared to those they obtain from the {\it Gaia} photometry and parallax, concluding the agreement is good, for DAs. They concluded the DA and DB and DBAs mean masses obtained from the {\it Gaia} data match within 2\%, but their Figs.~4, 5 and 12 show the disagreement for DBs, between spectroscopy and Gaia parallaxes is {\it larger} when the pure He 3D corrections are applied. As discussed in section~\ref{section:masses}, the introduction of the pure 3D correction for DBs is the cause of the reduction in the mean mass of DBs, and it is probably not real. \citet{Ourique19} show there is strong evidence for spectral evolution with effective temperature.

\citet{Latour18} analysed the hot subdwarfs of the globular cluster $\omega$ Cen, and found a ratio of 26\% sdBs ($T_\mathrm{eff} \leqslant 30,000$~K), 10\% sdOs ($T_\mathrm{eff} \geqslant 42,000$~K), and the majority as sdBs (intermediate $T_\mathrm{eff}$). They also found the majority of their sdOBs were helium-enriched, without a counterpart in Galactic field,
while we found 33/128=26\% of sdOBs are He-sdOBs.

\section{Conclusions}
We extended our search of white dwarf and subdwarf stars to SDSS DR14. 
In addition to searching all spectra with significant proper motion for new white dwarfs, we also
fitted known DAs and DBs that fell in our selection criteria. The SDSS flux calibration is based on hundreds of comparison stars and in general more accurate than those derived from single night observations.
We fitted the spectra of highest signal-to-noise for each star, taking into account that SDSS re-observes fields and improves the quality of the spectra.
Our classifications are independent from previous classifications, and should be considered improvements.

Of the total 37\,053 objects in our Table~\ref{tab:all.1},
only 6 per cent come from plates obtained after DR12, but only 13\,927 are in the SDSS DR~7 to DR~12 catalogues.
The DR7 to DR12 catalogues contain 35\,590 stars, including 29\,262 DAs, so our catalogues are not a subset or complete sets, but complementary. The total number of unique stars in \citet{dr7,dr10,dr12} and this DR14 catalogue is  52\,299, with 28\,681 DAs, 2287 DCs, 2148 DBs, 1126 DZs, 572 DQs, 137 DOs, 4 DS, 396 sdB, 410 sdOs, and 324 CVs.

For the first time we include 3D convection corrections to the derived effective temperatures of DBs, in addition to DAs. The obtained mean masses for DAs and DBs are lower than any previous determinations. For DAs, the main difference was the inclusion of more DAs cooler than $T_\mathrm{eff}=10\,000$~K, which show substantially smaller masses, while for DBs the inclusion of the convection correction was the main difference. \citet{Tremblay18} show the disagreement between the spectroscopic determinations and the Gaia parallaxes and colours increases when the 3D correction is applied.

The Gaia distances and colours show there is large spread in the region between cool white dwarfs and cool main sequence stars. Due to the considerable uncertainty in the parallax (of the same order of the parallax itself for most stars in this region), a reliable separation between different types of sdAs is not possible. This spread causes many stars to be in the region between the main sequence and the white dwarf cooling range, which is compatible with interacting binary evolution. This region is occupied by known sdBs, sdOs, CVs, and WD+MS binaries. The parallax uncertainties suggest most ($> 95$~per cent) of the sdAs in this intermediary reason are consistent with low-mass metal-poor halo stars, but a few could be products of binary evolution such as ELMs.

\section*{Acknowledgements}
This study was financed in part by the Coordena\c{c}\~ao de Aperfei\c{c}oamento de Pessoal de N\'{\i}vel Superior - Brasil (CAPES) - Finance Code 001, Conselho Nacional de Desenvolvimento Cient\'{\i}fico e Tecnol\'ogico - Brasil (CNPq), and Funda\c{c}\~ao de Amparo \`a Pesquisa do Rio Grande do Sul (FAPERGS) - Brasil. IP acknowledges support from the Deutsche Forschungsgemeinschaft under grant GE2506/12-1. Funding for the Sloan Digital Sky Survey IV has been provided by the Alfred P. Sloan Foundation, the U.S. Department of Energy Office of Science, and the Participating Institutions. SDSS-IV acknowledges
support and resources from the Center for High-Performance Computing at
the University of Utah. The SDSS web site is www.sdss.org.
SDSS-IV is managed by the Astrophysical Research Consortium for the 
Participating Institutions of the SDSS Collaboration including the 
Brazilian Participation Group, the Carnegie Institution for Science, 
Carnegie Mellon University, the Chilean Participation Group, the French Participation Group, Harvard-Smithsonian Center for Astrophysics, 
Instituto de Astrof\'{\i}sica de Canarias, The Johns Hopkins University, 
Kavli Institute for the Physics and Mathematics of the Universe (IPMU) / 
University of Tokyo, Lawrence Berkeley National Laboratory, 
Leibniz Institut f\"ur Astrophysik Potsdam (AIP),  
Max-Planck-Institut f\"ur Astronomie (MPIA Heidelberg), 
Max-Planck-Institut f\"ur Astrophysik (MPA Garching), 
Max-Planck-Institut f\"ur Extraterrestrische Physik (MPE), 
National Astronomical Observatories of China, New Mexico State University, 
New York University, University of Notre Dame, 
Observat\'ario Nacional / MCTI, The Ohio State University, 
Pennsylvania State University, Shanghai Astronomical Observatory, 
United Kingdom Participation Group,
Universidad Nacional Aut\'onoma de M\'exico, University of Arizona, 
University of Colorado Boulder, University of Oxford, University of Portsmouth, 
University of Utah, University of Virginia, University of Washington, University of Wisconsin, 
Vanderbilt University, and Yale University.

This research has made use of NASA's Astrophysics Data System Bibliographic Services,
SIMBAD database, operated at CDS, Strasbourg, France, and 
IRAF, distributed by the National Optical Astronomy Observatory, which is operated by the Association of Universities for Research in Astronomy (AURA) under a cooperative agreement with the National Science Foundation.
This work presents results from the European Space Agency (ESA) space mission Gaia. Gaia data are being processed by the Gaia Data Processing and Analysis Consortium (DPAC). Funding for the DPAC is provided by national institutions, in particular the institutions participating in the Gaia MultiLateral Agreement (MLA). The Gaia mission website is https://www.cosmos.esa.int/gaia. The Gaia archive website is https://archives.esac.esa.int/gaia.

The Gaia mission and data processing have financially been supported by, in alphabetical order by country:
the Algerian Centre de Recherche en Astronomie, Astrophysique et G\'eophysique of Bouzareah Observatory;
the Austrian Fonds zur F\"orderung der wissenschaftlichen Forschung (FWF) Hertha Firnberg Programme through grants T359, P20046, and P23737;
the BELgian federal Science Policy Office (BELSPO) through various PROgramme de D'eveloppement d'Exp\'eriences scientifiques (PRODEX) grants and the Polish Academy of Sciences - Fonds Wetenschappelijk Onderzoek through grant VS.091.16N;
the Brazil-France exchange programmes Funda\c{c}\~ao de Amparo \`a Pesquisa do Estado de S~ao Paulo (FAPESP) and Coordena\c{c}\~ao de Aperfeicoamento de Pessoal de N\'{\i}vel Superior (CAPES) - Comit\'e Fran\c{c}ais d'Evaluation de la Coop\'eration Universitaire et Scientifique avec le Br\'esil (COFECUB);
the Chilean Direcci\'on de Gesti\'on de la Investigaci\'on (DGI) at the University of Antofagasta and the Comit\'e Mixto ESO-Chile;
the National Science Foundation of China (NSFC) through grants 11573054 and 11703065;
the Czech-Republic Ministry of Education, Youth, and Sports through grant LG 15010, the Czech Space Office through ESA PECS contract 98058, and Charles University Prague through grant PRIMUS/SCI/17;
the Danish Ministry of Science;
the Estonian Ministry of Education and Research through grant IUT40-1;
the European Commission's Sixth Framework Programme through the European Leadership in Space Astrometry (ELSA) Marie Curie Research Training Network (MRTN-CT-2006-033481), through Marie Curie project PIOF-GA-2009-255267 (Space AsteroSeismology \& RR Lyrae stars, SAS-RRL), and through a Marie Curie Transfer-of-Knowledge (ToK) fellowship (MTKD-CT-2004-014188); the European Commission's Seventh Framework Programme through grant FP7-606740 (FP7-SPACE-2013-1) for the Gaia European Network for Improved data User Services (GENIUS) and through grant 264895 for the Gaia Research for European Astronomy Training (GREAT-ITN) network;
the European Research Council (ERC) through grants 320360 and 647208 and through the European Union's Horizon 2020 research and innovation programme through grants 670519 (Mixing and Angular Momentum tranSport of massIvE stars - MAMSIE) and 687378 (Small Bodies: Near and Far);
the European Science Foundation (ESF), in the framework of the Gaia Research for European Astronomy Training Research Network Programme (GREAT-ESF);
the European Space Agency (ESA) in the framework of the Gaia project, through the Plan for European Cooperating States (PECS) programme through grants for Slovenia, through contracts C98090 and 4000106398/12/NL/KML for Hungary, and through contract 4000115263/15/NL/IB for Germany;
the European Union (EU) through a European Regional Development Fund (ERDF) for Galicia, Spain;
the Academy of Finland and the Magnus Ehrnrooth Foundation;
the French Centre National de la Recherche Scientifique (CNRS) through action 'D\'efi MASTODONS' the Centre National d'Etudes Spatiales (CNES), the L'Agence Nationale de la Recherche (ANR)  'Investissements d'avenir' Initiatives D'EXcellence (IDEX) programme Paris Sciences et Lettres (PSL**) through grant ANR-10-IDEX-0001-02, the ANR D'efi de tous les savoirs' (DS10) programme through grant ANR-15-CE31-0007 for project 'Modelling the Milky Way in the Gaia era' (MOD4Gaia), the R\'egion Aquitaine, the Universit\'e de Bordeaux, and the Utinam Institute of the Universit\'e de Franche-Comt\'e, supported by the R\'egion de Franche-Comt\'e and the Institut des Sciences de l'Univers (INSU);
the German Aerospace Agency (Deutsches Zentrum f\"ur Luft- und Raumfahrt e.V., DLR) through grants 50QG0501, 50QG0601, 50QG0602, 50QG0701, 50QG0901, 50QG1001, 50QG1101, 50QG1401, 50QG1402, 50QG1403, and 50QG1404 and the Centre for Information Services and High Performance Computing (ZIH) at the Technische Universit\"at (TU) Dresden for generous allocations of computer time;
the Hungarian Academy of Sciences through the Lend\"ulet Programme LP2014-17 and the J\'anos Bolyai Research Scholarship (L. Moln\'ar and E. Plachy) and the Hungarian National Research, Development, and Innovation Office through grants NKFIH K-115709, PD-116175, and PD-121203;
the Science Foundation Ireland (SFI) through a Royal Society - SFI University Research Fellowship (M. Fraser);
the Israel Science Foundation (ISF) through grant 848/16;
the Agenzia Spaziale Italiana (ASI) through contracts I/037/08/0, I/058/10/0, 2014-025-R.0, and 2014-025-R.1.2015 to the Italian Istituto Nazionale di Astrofisica (INAF), contract 2014-049-R.0/1/2 to INAF dedicated to the Space Science Data Centre (SSDC, formerly known as the ASI Sciece Data Centre, ASDC), and contracts I/008/10/0, 2013/030/I.0, 2013-030-I.0.1-2015, and 2016-17-I.0 to the Aerospace Logistics Technology Engineering Company (ALTEC S.p.A.), and INAF;
the Netherlands Organisation for Scientific Research (NWO) through grant NWO-M-614.061.414 and through a VICI grant (A. Helmi) and the Netherlands Research School for Astronomy (NOVA);
the Polish National Science Centre through HARMONIA grant 2015/18/M/ST9/00544 and ETIUDA grants 2016/20/S/ST9/00162 and 2016/20/T/ST9/00170;
the Portugese Funda\c{c}\~ao para a Ci\^encia e a Tecnologia (FCT) through grant SFRH/BPD/74697/2010; the Strategic Programmes UID/FIS/00099/2013 for CENTRA and UID/EEA/00066/2013 for UNINOVA;
the Slovenian Research Agency through grant P1-0188;
the Spanish Ministry of Economy (MINECO/FEDER, UE) through grants ESP2014-55996-C2-1-R, ESP2014-55996-C2-2-R, ESP2016-80079-C2-1-R, and ESP2016-80079-C2-2-R, the Spanish Ministerio de Econom\'\i a, Industria y Competitividad through grant AyA2014-55216, the Spanish Ministerio de Educaci\'on, Cultura y Deporte (MECD) through grant FPU16/03827, the Institute of Cosmos Sciences University of Barcelona (ICCUB, Unidad de Excelencia 'Mar\'\i a de Maeztu') through grant MDM-2014-0369, the Xunta de Galicia and the Centros Singulares de Investigaci\'on de Galicia for the period 2016-2019 through the Centro de Investigaci\'on en Tecnolog\'{\i}as de la Informaci\'on y las Comunicaciones (CITIC), the Red Espa\~nola de Supercomputaci\'on (RES) computer resources at MareNostrum, and the Barcelona Supercomputing Centre - Centro Nacional de Supercomputaci\'on (BSC-CNS) through activities AECT-2016-1-0006, AECT-2016-2-0013, AECT-2016-3-0011, and AECT-2017-1-0020;
the Swedish National Space Board (SNSB/Rymdstyrelsen);
the Swiss State Secretariat for Education, Research, and Innovation through the ESA PRODEX programme, the Mesures d'Accompagnement, the Swiss Activit\'es Nationales Compl\'ementaires, and the Swiss National Science Foundation;
the United Kingdom Rutherford Appleton Laboratory, the United Kingdom Science and Technology Facilities Council (STFC) through grant ST/L006553/1, the United Kingdom Space Agency (UKSA) through grant ST/N000641/1 and ST/N001117/1, as well as a Particle Physics and Astronomy Research Council Grant PP/C503703/1.







\bsp	
\label{lastpage}
\end{document}